\documentclass[12pt]{article}
\usepackage{geometry}
\usepackage{amsmath, amssymb, graphicx,fullpage,color,mathtools,amsthm,xcolor}
\usepackage{caption, subcaption}
\usepackage{cite}
\usepackage[normalem]{ulem}

\usepackage{microtype}
\usepackage{hyperref,color}
\usepackage{diagbox}
\usepackage{algorithm}
\usepackage{algpseudocode}
\usepackage{commath}

\graphicspath{{Images/}}

\definecolor{webgreen}{rgb}{0,.35,0}
\definecolor{webbrown}{rgb}{.6,0,0}
\definecolor{RoyalBlue}{rgb}{0,0,0.9}
\definecolor{purp}{rgb}{0.6,0.05,0.8}
\definecolor{ora}{rgb}{0.7,0.35,0.02}

\hypersetup{
   colorlinks=true, linktocpage=true, 
   urlcolor=webbrown, linkcolor=RoyalBlue, citecolor=webgreen,
   pdfauthor={Salem Mosleh, Gary P. T. Choi, L. Mahadevan},
   pdfsubject={Data-driven quasiconformal morphodynamic flows}
}

\begin{document}

\author{Salem Mosleh$^{1,2,\dagger}$, Gary P. T. Choi$^{3,\dagger}$, L. Mahadevan$^{1,4,\ast}$\\
\\
\footnotesize{$^{1}$School of Engineering and Applied Sciences, Harvard University, Cambridge, MA, USA}\\
\footnotesize{$^{2}$Department of Natural Sciences, University of Maryland Eastern Shore, Princess Anne, MD, USA}\\
\footnotesize{$^{3}$Department of Mathematics, The Chinese University of Hong Kong, Hong Kong}\\
\footnotesize{$^{4}$Departments of Physics, and Organismic and Evolutionary Biology,}\\ 
\footnotesize{Harvard University, Cambridge, MA, USA}\\
\footnotesize{$^\dagger$S.M. and G.P.T.C. contributed equally to this work.}\\
\footnotesize{$^\ast$To whom correspondence should be addressed; E-mail: lmahadev@g.harvard.edu}
}
\title{Data-driven quasiconformal morphodynamic flows}
\date{}
\maketitle

\begin{abstract}

Temporal imaging of biological epithelial structures yields shape data at discrete time points, leading to a natural question: how can we reconstruct the most likely path of growth patterns consistent with these discrete observations? We present a physically plausible framework to solve this inverse problem by creating a framework that generalises quasiconformal maps to quasiconformal flows. By allowing for the spatio-temporal variation of the shear and dilatation fields during the growth process, subject to regulatory mechanisms, we are led to a type of generalised Ricci flow. When guided by observational data associated with surface shape as a function of time, this leads to a constrained optimization problem. Deploying our data-driven algorithmic approach to the shape of insect wings, leaves and even sculpted faces, we show how optimal quasiconformal flows allow us to characterise the morphogenesis of a range of surfaces. 

\end{abstract}

\section{Introduction}
 
Morphogenesis, the process by which organisms generate and regulate their shape, involves a complex interplay between biochemical and physical factors. A key goal is to identify the biophysical factors --- and how they interact with biochemical signaling --- that drive the local deformations and flows during morphogenesis \cite{thompson1917growth}. However, often we do not obtain the local growth and flow patterns of cells from experiments, which require live imaging. Thus a key mathematical challenge is to infer the local growth patterns from snapshots of growing structures taken at different developmental stages, possibly from different individuals, and learn the dynamical laws that generate these structures. This requires us to determine the mapping (or flow) rules that connect points on the different snapshots of the growing structure.

\begin{figure}[t]
 \includegraphics[width=\textwidth]{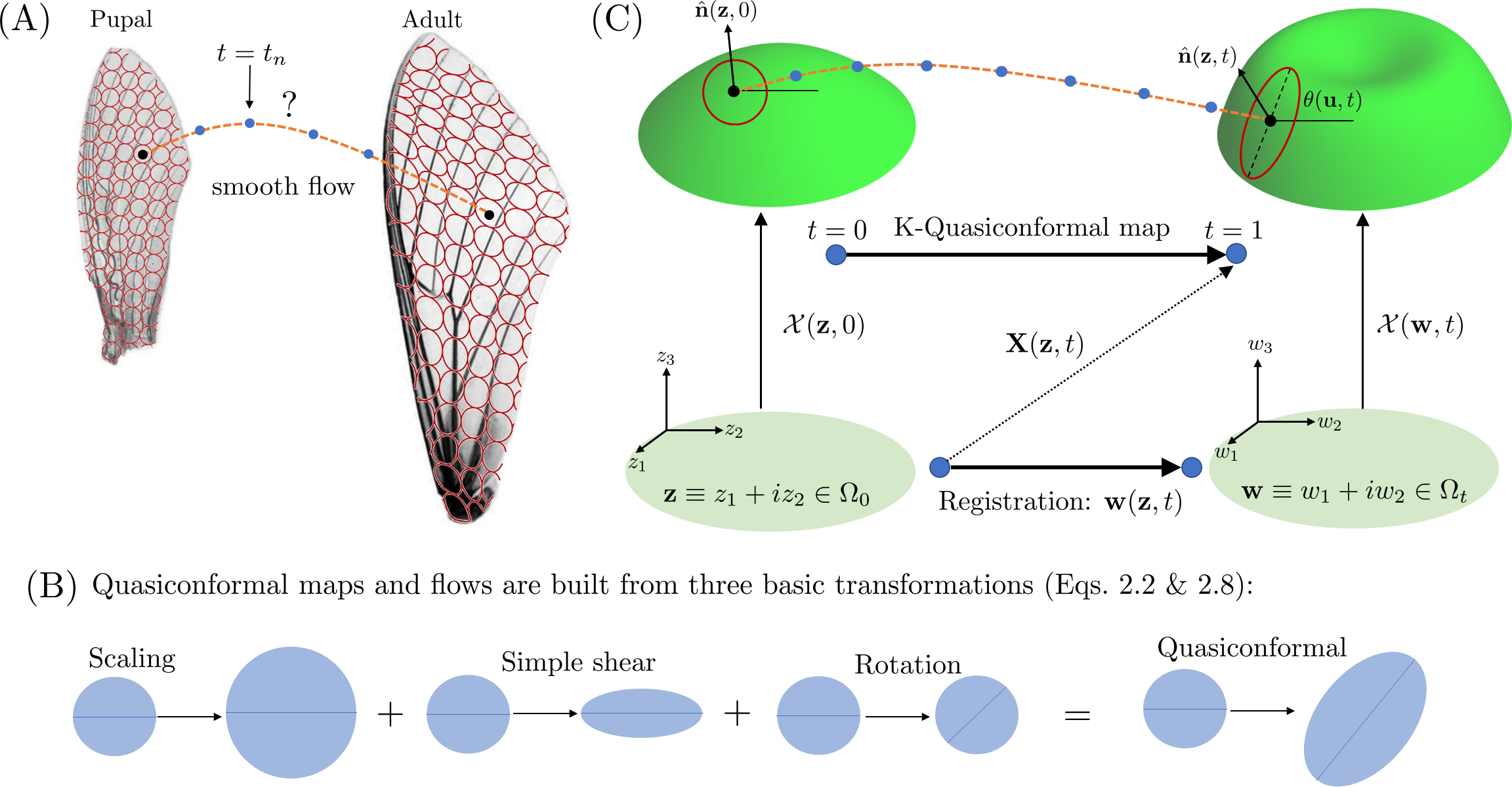}
 	\caption{\textbf{Quasi-conformal flows for morphodynamics.} (A) Images of growing structures do not provide the correspondence between points at different times $t$ (normalised by dividing with the total duration $T$). We need to infer the (Lagrangian) trajectory of points computationally. (B) Quasiconformal (QC) maps and flows are useful for accomplishing this, since they describe systems that deform smoothly with anisotropic growth, where infinitesimal circles deform into infinitesimal ellipses over time due to the action of scaling, shear, and rotation fields. (C) An illustration of the geometry of deforming surfaces. Each surface, for $t>0$, is given as the function $\mathcal{X}(\mathbf{w}, t):\Omega_t \in \mathbb{R}^2$. Our task is to find a suitable registration $\mathbf{w}(\mathbf{z}, t)$ that allows us to infer the true growth patterns shown in panel (A) by constructing the function $\mathbf{X}(\mathbf{z}, t) \equiv \mathcal{X}(\mathbf{w}(\mathbf{z}, t), t)$. Here $\hat{\mathbf{n}}(\mathbf{z}, t)$ is the unit normal to the surface at time $t$.} 
  \label{fig:geometry} 
\end{figure}

One way to address this correspondence problem is through the use of conformal maps between planar domains or two-dimensional surfaces embedded in three dimensions. These maps do not distort angles~\cite{ahlfors1953complex, mumford-conformal, levy-conformal, desburn-conformal}. For example given two simply connected regions in the plane, by the Riemann mapping theorem, a unique (up to M\"{o}bius transformations) conformal map exists between them and can be used to determine the correspondence between all points in the interior domains given their boundary curves. This mathematical fact has been put to use to study growing leaves~\cite{mitchison2016conformal}, insect wings \cite{alba2021global}, and other biological systems \cite{gu2004genus,choi2015flash}. However conformal maps --- which generate isotropic growth, mapping infinitesimal circles to rescaled infinitesimal circles --- are inadequate to describe biological morphogenesis where anisotropic growth --- which distorts angles and maps infinitesimal circles to infinitesimal ellipses (Fig.~\ref{fig:geometry}A) --- is the rule rather than the exception \cite{kierzkowski2019growth, zhao2020microtubule}. 

For nearly a century, a generalisation of conformal maps known as quasi-conformal maps that do account for dilatation, shear and rotation (Fig.~\ref{fig:geometry}B) have been known \cite{ahlfors2006lectures}, although it is only recently that they have been deployed to study morphogenesis, e.g. to allow for planar deformations that minimise elastic distortions while also accounting for the presence of landmarks (points whose mapping is given \textit{a priori})~\cite{jones2013planar,choi2018planar}. When combined with different types of regularizations, e.g. minimising squared gradients of the Beltrami fields \cite{jones2013planar} or restrictions to smooth Teichm\"uller maps \cite{choi2018planar}, it has been possible to study both the phylogeny and ontogeny of wing shapes using 2D quasiconformal maps. In parallel, the field of computational quasi-conformal mappings has grown rapidly with applications to complex 3D shapes such as teeth, faces etc.~\cite{choi2020tooth,choi2020shape}.

%Maps are not sufficient. We need flows
As we gain the ability to continuously monitor growing systems, and extract their shape over time, it is natural to ask whether one can take advantage of the temporal structure of the growth process, by considering \textit{flows} (instead of maps between discrete times points), which give smooth transformations of a shape into another shape over time \cite{klassen2004analysis, beg2005computing}. To this end, we generalise $K$-quasiconformal maps, which deform (infinitesimal) circles to ellipses whose eccentricity is bounded by the value $K$, to $k$-quasiconformal \textit{flows} with time duration $T$, defined as the $dt \to 0$ sequence of ($k dt$)-quasiconformal maps. Considering $k$-quasiconformal flows leads to interesting new mathematical problems, for example, a Teichm{\"u}ller map (a $K$-quasiconformal map with minimal $K$) between two surfaces may not be the same as the $k$-quasiconformal flow, with minimal $k$, connecting them. 

As mentioned above, the prescribed data often is not sufficient (due to incomplete tracking of points) to uniquely determine a quasi-conformal map and, therefore, additional criteria are required. Common criteria are minimal distortion \cite{srivastava2016functional} and minimal spatial variation \cite{jones2013planar} mappings. However, growing structures (including leaves and electrochemical interfaces) are governed by dynamical processes that couple the curvature, mechanical strain, and other fields. Here, in addition to considering minimal distortion and spatial variation flow, we find optimal quasi-conformal flows that fit the prescribed data to a dynamical equation that takes the form of a geometric partial differential equation, similar to a generalised Ricci flow. Our approach is a step towards blending data-driven and physics-based approaches since it allows us to automate the process of the discovery of growth laws and predict growth patterns from incomplete data.  

This paper is organised as follows. In Section \ref{sec:theory} we discuss the geometry and kinematics relevant to evolving surfaces and define the cost function that will be minimised in the following sections. Section \ref{sec:discrete} will discuss the discretisation and numerical implementation, and Section \ref{sec:results} will present experiments performed on various natural and synthetic examples of flowing surfaces. We conclude our work and discuss future directions in Section~\ref{sect:conclusion}.

\section{Formulation of quasiconformal flow models} \label{sec:theory}

This section introduces the formalism for optimal quasiconformal flows and defines the cost function that we use in the following sections.

\subsection{Geometry of surface and quasiconformal maps}

We start with a smooth one parameter family (flow) of surfaces $\tilde{\mathcal{X}}(\mathbf{w}, \tilde{t})$, which are embedded in $\mathbb{R}^3$ with parametrization $\mathbf{w} \in \Omega_t \subset \mathbb{R}^2$ and time $\tilde{t} \in [0, T]$ (see Fig.~\ref{fig:geometry}C). We normalise lengths so that $\mathcal{X}(\mathbf{w}, t) \equiv \tilde{\mathcal{X}}(\mathbf{w}, t)/\sqrt{\mathcal{A}}$, where $\mathcal{A}$ is the area of $\tilde{\mathcal{X}}(\mathbf{w}, 0)$, and normalise time so that $t \equiv \tilde{t}/T \in [0, 1]$. Our goal is to find a registration --- a correspondence between points on the surfaces across time --- as a coordinate system $\mathbf{z} = z_1 + i z_2 \in \Omega_0 \subset \mathbb{R}^2$ and a family of maps $\mathbf{w}(\mathbf{z}, t)$ that gives a common parameterisation of the surface across time, $\mathbf{X}(\mathbf{z}, t) \equiv \mathcal{X}(\mathbf{w}(\mathbf{z}, t), t): \Omega_0 \to \mathbb{R}^3$. Thus, for a fixed $\mathbf{z}$, the function $\mathbf{X}(\mathbf{z}, t)$ gives the trajectory of the ``same'' point evolving over time (orange dashed curve in Fig.~\ref{fig:geometry}C) and our task is to obtain these trajectories.

For notational convenience, we will express the coordinate vector as $\mathbf{z} \equiv z_i$ ($i = 1, 2$), and define derivative operators as $\partial_i \equiv \partial/\partial z_i$. Furthermore, we will use the Einstein summation convention, where pairs of repeated indices in an expression are summed over. Using this notation and the chain rule, a tangent vector to the surface that connects two infinitesimally separated points $z_i$ and $z_i + dz_i$ at time $t$ can be written as $d \mathbf{X}(\mathbf{z}, t) =  \partial_i \mathbf{X}(\mathbf{z}, t) \;dz_i$.

To describe the kinematics of growing surfaces, it is useful to consider how angles and lengths change as the surface evolves. The length squared of the line segment connecting the points $z_i$ and $z_i + dz_i$ is given by 
\begin{eqnarray}
    d\ell^2(t) = |d \mathbf{X}(\mathbf{z}, t)|^2 = g_{ij}(\mathbf{z}, t) \;dz_i dz_j, \;\;\;\;\;\;\;  g_{ij}(\mathbf{z}, t) \equiv \partial_i \mathbf{X}(\mathbf{z}, t) \cdot \partial_j \mathbf{X}(\mathbf{z}, t), \label{eq:metric}
\end{eqnarray}
where $g_{ij}(\mathbf{z}, t)$ gives the components of the metric tensor in the $z_i$ coordinate system. Using the metric tensor, we can define lengths of (and therefore angles between) line segments on the surface at time $t$. The inverse of the metric is denoted (with upper indices) as $g^{ij}(\mathbf{z}, t)$, and satisfies $g^{ij}(\mathbf{z}, t) g_{jk}(\mathbf{z}, t) = \delta_{ik}$, where $\delta_{ik}$ is the Kronecker delta. 

Next, we describe the metric changes in terms of rotations, dilations, and shear of the tangent plane. To facilitate this, we consider the Cartesian coordinates on the plane tangent to the surface $\mathbf{X}(\mathbf{z}, 0)$ at $\mathbf{z}$, which we denote as $\tilde{\mathbf{z}}$. The Jacobian of the transformation between $\tilde{\mathbf{z}}$ and $\mathbf{z}$ is given by $\mathbf{J}_{i}(\mathbf{z}) \equiv \partial_i \tilde{\mathbf{z}}(\mathbf{z})$ and the initial metric is given by ${g}_{ij}(\mathbf{z}, 0) = \mathbf{J}_i^\top(\mathbf{z}) \cdot \mathbf{J}_j(\mathbf{z})$, where the superscript $^\top$ indicates a vector or matrix transpose. As the surface evolves over time, the tangent plane at $\mathbf{X}(\mathbf{z}, 0)$ will be convected by the flow and will undergo rotation, dilation, and shear that we explicitly account for using a polar decomposition of the deformation gradient of the tangent plane followed by a coordinate transform --- using the Jacobian $\mathbf{J}_i(\mathbf{z})$ --- which leads to the following expression for the metric tensor at time $t$,
\begin{eqnarray}
{g}_{ij}(\mathbf{z}, t) = \omega(\mathbf{z}, t) \mathbf{J}^\top_{i}(\mathbf{z}) \mathbf{R}^\top(\mathbf{z} ,t) \mathbf{E}(\mathbf{z} ,t) \mathbf{R}(\mathbf{z} ,t) \mathbf{J}_j(\mathbf{z}), \;\;
\mathbf{E}(\mathbf{z} ,t) \equiv \begin{pmatrix}
1 + \epsilon(\mathbf{z}, t) & 0 \\
0 & \frac{1}{1 + \epsilon(\mathbf{z}, t)}
\end{pmatrix},\label{eq:metric-decomp}
\end{eqnarray}
where the scalar $\omega(\mathbf{z} ,t) > 0$ is the conformal factor giving the dilation of the transformation, $\epsilon(\mathbf{z}, t) \geq 0$ relates to the eccentricity of an ellipse at time $t$ that started as an infinitesimal circle on $\mathbf{X}(\mathbf{z},0)$, with $1 + \epsilon(\mathbf{z}, t)$ being the ratio of major to minor axes and transformations for which $1 + \epsilon(\mathbf{z}, t) \leq K$ would be called $K$-quasiconformal. The 2D rotation matrix $\mathbf{R}(\mathbf{z} ,t)$ gives the direction of the major axes of the ellipse in the tangent plane at $\mathbf{z}$ (see Fig.~\ref{fig:geometry}). The Beltrami coefficient \cite{ahlfors2006lectures} of this transformation is then defined as 
\begin{eqnarray}
    \mu(\mathbf{z}, t)  = \frac{\epsilon(\mathbf{z}, t)}{2 + \epsilon(\mathbf{z}, t) } \exp\left[- 2  i \theta(\mathbf{z} ,t) \right],
\end{eqnarray}
where $\theta(\mathbf{z},t)$ is the rotation angle defining the matrix $\mathbf{R}(\mathbf{z},t)$.

In addition to length, the curvature of line segments on the surface will be a useful measure of its geometry. For a curve on the surface tangent to the line segment connecting the points $z_i$ and $z_i + dz_i$, whose length is given in Eq.~\eqref{eq:metric}, the normal curvature is defined as the projection of the curvature of the curve in space $(\boldsymbol{\kappa})$ in the direction of the unit normal to the surface $\hat{\mathbf{n}}(\mathbf{z}, t)$. Thus, we write
\begin{eqnarray}
    \kappa_N \equiv \boldsymbol{\kappa} \cdot \hat{\mathbf{n}}(\mathbf{z}, t) = b_{ij}(\mathbf{z}, t) \frac{dz_i}{d\ell} \frac{dz_j}{d\ell}, \;\;\;\;\;\;\; b_{ij}(\mathbf{z}, t) \equiv \hat{\mathbf{n}}(\mathbf{z}, t)\cdot \partial_k \partial_j \mathbf{X}(\mathbf{z}, t),
\end{eqnarray}
 where we defined the curvature tensor $b_{ij}(\mathbf{z}, t)$. At a given point on the surface, the principal directions (directions of maximum and minimum normal curvatures) will be denoted as $\hat{\boldsymbol{\kappa}}_1(\mathbf{z}, t)$ and $\hat{\boldsymbol{\kappa}}_2(\mathbf{z}, t)$, with corresponding principal curvatures ${\kappa}_1(\mathbf{z}, t)$ and ${\kappa}_2(\mathbf{z}, t)$. From the principal curvatures we can define the mean curvature $H(\mathbf{z}, t) \equiv [\kappa_1(\mathbf{z}, t) + \kappa_2(\mathbf{z}, t)]/2$ and Gaussian curvature $K(\mathbf{z}, t) \equiv \kappa_1(\mathbf{z}, t)  \kappa_2(\mathbf{z}, t)$.

\subsection{Quasiconformal flows}
In this section, we consider the temporal rate of change of the quantities defined in the previous section. Since we assume that the trajectories of points $\mathbf{X}(\mathbf{z}, t)$ are smooth in time, we define the velocity field as
\begin{eqnarray}
   \mathbf{V}(\mathbf{z}, t) \equiv \dot{\mathbf{X}}(\mathbf{z}, t) = \mathbf{V}_{\parallel}(\mathbf{z}, t) + V_{\perp}(\mathbf{z}, t) \;\hat{\mathbf{n}}(\mathbf{z}, t),  \;\;\;\; \mathbf{z} \in \Omega_0,\label{eq:velocity}
\end{eqnarray}
where the over dot denotes a partial derivative with respect to time, and we have split the velocity into a component that is parallel to the surface, and a component along the surface normal $\hat{\mathbf{n}}(\mathbf{z}, t)$. For points along the boundary of an open surface, we define the unit tangent vector to the boundary as $\hat{\boldsymbol{\tau}}(\mathbf{z}, t)$ and write
\begin{eqnarray}
    \mathbf{V}_{\parallel}(\mathbf{z}, t) = V_\tau(\mathbf{z}, t) \hat{\boldsymbol{\tau}}(\mathbf{z}, t) + V_\partial(\mathbf{z}, t) \hat{\boldsymbol{n}}_\partial(\mathbf{z}, t), \;\;\; \hat{\boldsymbol{n}}_\partial(\mathbf{z}, t)\equiv  \hat{\boldsymbol{\tau}}(\mathbf{z}, t)\times\hat{\mathbf{n}}(\mathbf{z}, t), \;\;\mathbf{z} \in \partial\Omega_0. \label{eq:boundary-vel}
\end{eqnarray}
The components of the velocity normal to the surface and normal to the boundary curve, $V_\perp(\mathbf{z}, t)$ and $V_\partial(\mathbf{z}, t)$, will be independent of registration since changing the registration $\mathbf{w}(\mathbf{z}, t)$ at time $t$ can only move the points parallel to the surface $\mathbf{X}(\mathbf{z}, t)$ \cite{tumpach2015gauge}. Therefore, the task of finding the registration $\mathbf{w}(\mathbf{z}, t)$ given a flow of surfaces $\mathcal{X}(\mathbf{w}, t)$ is equivalent to finding  $\mathbf{V}_{\parallel}(\mathbf{z}, t)$ given the perpendicular velocities $V_{\perp}(\mathbf{z}, t)$ and $V_{\partial}(\mathbf{z}, t)$. 

Next, we consider the rate of change of the metric, or strain rate tensor, which can be found by taking the time derivative of Eq.~\eqref{eq:metric}, 
\begin{eqnarray}
     \dot{g}_{ij} (\mathbf{z}, t) = \partial_i \mathbf{X}(\mathbf{z}, t) \cdot \partial_j \mathbf{V}(\mathbf{z}, t) + \partial_i \mathbf{V}(\mathbf{z}, t) \cdot \partial_j \mathbf{X}(\mathbf{z}, t). \label{eq:strain-rate}
\end{eqnarray}
Note that, for a fixed registration, the coordinate $\mathbf{z}$ represents a Lagrangian (material) coordinate system and therefore the time derivative of the metric tensor is still a tensor with the same rank. The rate of change of the metric may also be expressed using the decomposition given in Eq.~\eqref{eq:metric-decomp},
\begin{eqnarray}
\dot{g}_{ij}  = \frac{\dot{\omega}}{\omega} \;g_{ij} + \frac{\dot{\epsilon}\; \omega}{1 + \epsilon} \mathbf{J}^\top_{i} \;\mathbf{R}^\top\; \mathbf{E}\; \sigma_Z \;\mathbf{R}  \;\mathbf{J}_j  +
\omega \;\mathbf{J}^\top_{i}  \left[\dot{\mathbf{R}}^\top\; \mathbf{R}, \mathbf{R}^\top \;\mathbf{E} \; \mathbf{R} \right]  \mathbf{J}_j , \;\;\sigma_Z = \begin{pmatrix}
            1 & 0 \\
            0 & -1
    \end{pmatrix}, \label{eq:strain-rate-decomp}
\end{eqnarray} 
where $\left[\mathcal{O}_1, \mathcal{O}_2\right] \equiv \mathcal{O}_1 \mathcal{O}_2 - \mathcal{O}_2 \mathcal{O}_1$, for matrices $\mathcal{O}_1$ and $\mathcal{O}_2$. The three terms in this expression represent changes in the metric due to dilations, shears, and rotations respectively. Equation~\eqref{eq:strain-rate-decomp} gives the dilation and shear rates as: 
\begin{eqnarray}
&~&\mathcal{D}^2(\mathbf{z}, t) \equiv \frac{1}{2} Tr\left[\dot{g}_{ij} (\mathbf{z}, t)\right]^2 \equiv \frac{1}{2} \left(g^{ij}(\mathbf{z}, t) \dot{g}_{ij} (\mathbf{z}, t)\right)^2 =  \left(\frac{\dot{\omega}(\mathbf{z}, t)}{\omega(\mathbf{z}, t)}\right)^2, \label{eq:dilation}\\ 
&~&\mathcal{S}^2(\mathbf{z}, t)\equiv \frac{1}{2} Tr\left[\{\dot{g}_{ij} (\mathbf{z}, t)\}^2\right] =
 \left(\frac{\dot{\epsilon}(\mathbf{z}, t)}{1 + \epsilon(\mathbf{z}, t)}\right)^2 +  \frac{\left[2 + \epsilon(\mathbf{z}, t)\right]^2 \; \epsilon^2(\mathbf{z}, t) \; \dot{\theta}^2(\mathbf{z}, t)}{\left[1 + \epsilon(\mathbf{z}, t)\right]^2},\;\;\;\;\;\;\;\;\; \label{eq:shear}
\end{eqnarray} 
where $\{\dot{g}_{ij} (\mathbf{z}, t)\}$ is the traceless part of the strain rate tensor. Note that the quantities $\mathcal{D}(\mathbf{z}, t)$ and $\mathcal{S}(\mathbf{z}, t)$ are instantaneous quantities, depending only on the change in the metric at time $t$, whereas $\omega(\mathbf{z}, t), \epsilon(\mathbf{z}, t)$ and $\theta(\mathbf{z}, t)$ depend on the initial surface, with $\omega(\mathbf{z}, 0) = 1$ and $\epsilon(\mathbf{z}, 0) = 0$.

The quantities described so far in this section are intrinsic (depending on the metric) and do not quantify changes in embedding. For example, when unrolling a cylinder into a flat sheet, the strain rate tensor would vanish, $\dot{g}_{ij}(\mathbf{z}, t) = 0$. To account for changes in embedding that do not stretch the surface (isometric deformations), we follow Ref.~\cite{jermyn2012elastic} and define the bending strain as the rate of change of the unit normal, $\dot{\mathbf{n}}(\mathbf{z}, t) \equiv \partial \hat{\mathbf{n}}(\mathbf{z}, t)/\partial t$, to write a bending-like term in the cost function as
\begin{eqnarray}
    \mathcal{C}_{bend}\left[\mathbf{w}(\mathbf{z}, t)\right] =  A_3 \int dt   d\mathcal{A}\;   \;\dot{\mathbf{n}}^2(\mathbf{z}, t), \label{eq:bending-cost}
\end{eqnarray}
where the area element is defined using the determinant of the metric, $g(\mathbf{z},t)$, as $d\mathcal{A} \equiv dz_1 dz_2 \sqrt{g(\mathbf{z},t)}$ and $A_{3}$ is a dimensionless constant.

\subsection{Flow rules}

Next, we consider the criteria by which a quasiconformal flow is uniquely selected from the many possible ones that fit the given data $\mathcal{X}(\mathbf{w}, t)$. In addition to common choices that focus on the parsimony of the mapping --- such as minimising an energy that measures the extent of the distortion and its spatial variation --- we incorporate dynamical processes that generate and regulate morphogenetic flows into our framework. For concreteness, we consider shapes that may be generated from a combination of Ricci flows \cite{hamilton1988ricci} and mean curvature flows \cite{mullins1956}, which may be relevant for growing leaves, beaks, and bacterial cell shapes~\cite{kc-review, al2022feedback,al2022grow, al2021geometry}. To accomplish this, we define a growth strain tensor which is the difference between the strain rate tensor $\dot{g}_{ij}$ and the predicted value based on the dynamical law, which can be written as an expansion involving terms proportional to the metric and curvature tensors \cite{al2018growth},
\begin{eqnarray}
    G_{ij}(\mathbf{z}, t) = \frac{1}{2}\dot{g}_{ij}(\mathbf{z}, t) - \lambda_1 \;g_{ij}(\mathbf{z}, t) - \lambda_2 \;  H(\mathbf{z}, t) b_{ij}(\mathbf{z}, t) - \lambda_3  \; K(\mathbf{z}, t) g_{ij}(\mathbf{z}, t), \label{eq:growth-strain}
\end{eqnarray}
where the dimensionless parameters $\lambda_{\boldsymbol{\star}}$ are independent of $\mathbf{z}$ but may depend on time. If the flow $\mathcal{X}(\mathbf{w}, z)$ is in fact generated by the dynamical law, there exists a registration $\mathbf{w}(\mathbf{z}, t)$ such that $G_{ij}(\mathbf{z}, t) = 0$. For such flows, the $\lambda_1$ term gives exponential rate of expansion or contraction, the $\lambda_2$ term is related to the mean curvature flow, while the $\lambda_3$ term is related to the Ricci flow \cite{al2018growth}.

To find a quasiconformal flow that is as close to satisfying $G_{ij}(\mathbf{z}, t) = 0$ as possible, we will add the following terms to the cost function
\begin{eqnarray}    \mathcal{C}_{viscous}\left[\mathbf{w}(\mathbf{z}, t)\right] = \int dt d\mathcal{A}  \;\mathcal{K}_1^{ijkl} \;G_{ij}\; G_{kl}(\mathbf{z}, t) + C_g \int dt  \left[{\lambda}^2_1(t) + {\lambda}^2_2(t) + {\lambda}^2_3(t)\right], \label{eq:viscous-cost}
\end{eqnarray}
where $C_g$ is a dimensionless regularising constant and $\mathcal{K}_1^{ijkl}$ is an isotropic rigidity matrix, which we parameterise as,
\begin{eqnarray}
    \mathcal{K}_{1}^{ijkl}(\mathbf{z}, t) \equiv (A_{1} - B_{1}/2) g^{ij}(\mathbf{z}, t)  g^{kl}(\mathbf{z}, t) + B_{1} \; g^{ik}(\mathbf{z}, t)  g^{jl}(\mathbf{z}, t), \label{eq:rigidity-mat1}
\end{eqnarray}
where $A_1$ (analogous to bulk modulus) and $B_1$ (shear modulus) are dimensionless constants. Note that when setting $\lambda_1 = \lambda_2 = \lambda_3 = 0$, Eq.~\eqref{eq:viscous-cost} measures the extent of the distortion, since $G_{ij}(\mathbf{z}, t)$ reduces to the strain rate $\dot{g}_{ij}(\mathbf{z}, t)/2$ in that case.

Furthermore, to obtain best-fitting quasiconformal flows that are as smooth as possible, we will add the following term (analogous to a Dirichlet energy \cite{vector-Dirichlet-2020}) to the cost 
\begin{eqnarray}
    \mathcal{C}_{grad}\left[\mathbf{w}(\mathbf{z}, t)\right] =  \int dt d\mathcal{A} \; \mathcal{K}_2^{ijkl}(\mathbf{z}, t)  \;\nabla_m G_{ij}(\mathbf{z}, t) \nabla^m G_{kl}(\mathbf{z}, t),  \label{eq:grad-cost}
\end{eqnarray}
 where $\nabla_m$ is the covariant derivative \cite{do2016differential}, and $\mathcal{K}_2^{ijkl}$ is an isotropic rigidity matrix defined as,
\begin{eqnarray}
    \mathcal{K}_{2}^{ijkl}(\mathbf{z}, t) \equiv (A_{2} - B_{2}/2) g^{ij}(\mathbf{z}, t)  g^{kl}(\mathbf{z}, t) + B_{2} \; g^{ik}(\mathbf{z}, t)  g^{jl}(\mathbf{z}, t), \label{eq:rigidity-mat2}
\end{eqnarray}
where $A_2$ and $B_2$ are constants.

\subsection{Imposing (soft) constraints}
 Since our goal is to find an optimal quasiconformal flow that coincides with the given data $\mathcal{X}(\mathbf{w}, t)$, we need to impose constraints that fix the computed flow to the data. As mentioned after Eq.~\eqref{eq:boundary-vel}, this amounts to fixing the components of the velocity field that are normal to the surface and its boundary curve, while allowing the tangent components of the velocity field to vary. We constrain the normal components by adding the following terms to the cost function being minimised: 
\begin{eqnarray}
   \mathcal{C}_{normal}\left[\mathbf{w}(\mathbf{z}, t)\right] =  C_n \int_{\mathcal{X}} dt d\mathcal{A}\left[V_\perp(\mathbf{z}, t) - \bar{V}_\perp(\mathbf{z}, t)\right]^2,  \label{eq:normal-cost}\\ \mathcal{C}_{boundary}\left[\mathbf{w}(\mathbf{z}, t)\right] = C_b \int_{\partial\mathcal{X}} dt ds \left[V_\partial(\mathbf{z}, t) - \bar{V}_\partial(\mathbf{z}, t)\right]^2, \label{eq:boundary-cost}
\end{eqnarray}
where $\bar{V}_\perp(\mathbf{z}, t)$ and $\bar{V}_{\partial}(\mathbf{z}, t)$ are the registration independent normal displacements that are prescribed by the given data, and $C_n, C_b$ are dimensionless constants. Note that the first integral is over the area of the surface, while the second is over its boundary curve.

Landmarks are points for which we know the trajectory \textit{a priori}, either from experiments or other considerations \cite{webster2010practical}. If we take $\mathbf{z}_\star$ as the coordinate of a given landmark point, whose velocity is given by $\bar{\mathbf{V}}_\star(t)$, we account for landmark constraints by adding the following to the cost function: 
\begin{eqnarray}  
\mathcal{C}_{landmark}\left[\mathbf{w}(\mathbf{z}, t)\right] =  C_L \int dt \sum_{\star}  \left[\mathbf{V}(\mathbf{z}_\star, t) - \bar{\mathbf{V}}_\star(t)\right]^2, \;\;\;\label{eq:landmark-cost}
\end{eqnarray}
where $C_L$ is a dimensionless constant and we sum over all landmarks.

\begin{table}[t]
    \centering
\resizebox{\textwidth}{!}{
\begin{tabular}{|c|c|}\hline
\bf Terms & \bf Equations\\ \hline
Growth strain [Eq.~\eqref{eq:growth-strain}] & $G_{ij}(\mathbf{z}, t) = \frac{1}{2}\dot{g}_{ij}(\mathbf{z}, t) - \lambda_1 \;g_{ij}(\mathbf{z}, t) - \lambda_2 \;  H(\mathbf{z}, t) b_{ij}(\mathbf{z}, t) - \lambda_3  \; K(\mathbf{z}, t) g_{ij}(\mathbf{z}, t)$ \\ \hline
Viscous cost [Eq.~\eqref{eq:viscous-cost}] & $\mathcal{C}_{viscous}\left[ \mathbf{w}(\mathbf{z}, t)\right] = \int dt d\mathcal{A}  \left( \left(A_1 - \frac{B_1}{2}\right) \;Tr\left[G_{ij}\right]^2  + B_1 Tr\left[G_{ij}^2\right] \right) + C_g \int dt  \sum_{s=1}^3 \lambda_s^2(t)$ \\ \hline
Spatial gradient cost [Eq.~\eqref{eq:grad-cost}] & $\mathcal{C}_{grad}\left[\mathbf{w}(\mathbf{z}, t)\right] = \int dt d\mathcal{A}  \left( \left(A_2 - \frac{B_2}{2}\right) \;Tr\left[{\nabla}_k G_{ij}\right]^2  + B_2 Tr\left[{\nabla}_k G_{ij}^2\right] \right)$\\ \hline
Bending cost [Eq.~\eqref{eq:bending-cost}] & $\mathcal{C}_{bend}\left[\mathbf{w}(\mathbf{z}, t)\right] =  A_3 \int dt   d\mathcal{A}\;   \;\dot{\mathbf{n}}^2(\mathbf{z}, t) $\\ \hline
Enforcing normal displacement [Eq.~\eqref{eq:normal-cost}] & $\mathcal{C}_{normal}\left[\mathbf{w}(\mathbf{z}, t)\right] =  C_n \int_{\mathcal{X}_t} dt d\mathcal{A}\left[V_\perp(\mathbf{z}, t) - \bar{V}_\perp(\mathbf{z}, t)\right]^2$\\ \hline
Enforcing boundary displacements [Eq.~\eqref{eq:boundary-cost}] & $\mathcal{C}_{boundary}\left[\mathbf{w}(\mathbf{z}, t)\right] = C_b \int_{\partial\mathcal{X}_t} dt ds \left[V_\partial(\mathbf{z}, t) - \bar{V}_\partial(\mathbf{z}, t)\right]^2$\\ \hline
Enforcing landmark velocities [Eq.~\eqref{eq:landmark-cost}] & $\mathcal{C}_{landmark}\left[\mathbf{w}(\mathbf{z}, t)\right] =  C_L \int dt \sum_{\star}  \left[\mathbf{V}(\mathbf{z}_\star, t) - \bar{\mathbf{V}}_\star(t)\right]^2$\\ \hline
Overall cost function [Eq.~\eqref{eq:total-cost-function}] & $\mathcal{C}\left[\mathbf{w}(\mathbf{z}, t)\right] = \mathcal{C}_{viscous} + \mathcal{C}_{grad} + \mathcal{C}_{bend}  + \mathcal{C}_{normal} + \mathcal{C}_{boundary} + \mathcal{C}_{landmark}$\\ \hline
\end{tabular}
}
 \caption{\textbf{The terms and equations in the models for quasiconformal flows.}  The growth strain is the difference between the strain rate tensor $\dot{g}_{ij}(\mathbf{z}, t)$ and the value predicted from the geometric flow. $\mathcal{C}_{viscous}$ penalises non-zero growth strain tensor, while $\mathcal{C}_{grad}$ penalises spatial variations. The bending energy $\mathcal{C}_{bend}$ penalises changes in the shape that do not stretch the surface and the last three costs enforce the constraints from data. In practice, the constraints are enforced by choosing $C_n = C_b = C_L = 10^5$.}
  \label{tab:runs} 
\end{table} 

We are now in a position to define the cost function that will be used for the rest of the paper to be: 
\begin{eqnarray}
    &\mathcal{C}_{total}\left[\mathbf{w}(\mathbf{z}, t)\right] = \mathcal{C}_{viscous} + \mathcal{C}_{grad} + \mathcal{C}_{bend}  + \mathcal{C}_{normal} + \mathcal{C}_{boundary} + \mathcal{C}_{landmark}, \;\;\;\;\;\;\label{eq:total-cost-function} 
\end{eqnarray}
 where the individual terms are defined in Eqs.~(\ref{eq:bending-cost}--\ref{eq:grad-cost}) and the dimensionless parameters we choose are $A_1, B_1, A_2, A_3$, and the constraints will be enforced by choosing $C_n = C_b = C_L = 10^5$ for all models (see Table~\ref{tab:runs} for a summary of the terms and equations). 
 
 We will consider four different choices of parameters in this paper (see Table~\ref{tab:parameter}): 
 \begin{itemize}
     \item Almost-Conformal ($A_1 = 1$, all other parameters zero), which generates an as-conformal-as-possible fit to the flow;
     \item Viscous ($A_1 = B_1 = A_3 = 1$, and all other parameters zero), which generates a minimal distortion fit to the flow; 
     \item Almost-Uniform ($A_2 = B_2 = 1$, and all other parameters zero), which minimises the variation of the flow in space; 
     \item Geometric ($A_1 = B_1 = 1, C_g = 0.1,$ solve for $\lambda_1, \lambda_2, \lambda_3 \neq 0$, and all other parameters zero), which fits the flow to a dynamical equation, which in this paper is a geometric flow parameterised by $\lambda_1, \lambda_2$, and $\lambda_3$ (which are solved for during our fitting procedure).
 \end{itemize}

\begin{table}[t]
    \centering
\footnotesize{
\begin{tabular}{|c|c|c|c|c|c|c|}\hline
\backslashbox{\bf Models}{\bf Parameters} & $A_1$  & $B_1$  & $A_2$ & $B_2$ & $A_3$  & $\lambda_1, \lambda_2, \lambda_3$\\ \hline
Almost-Conformal & 1 & 0 & 0 & 0 & 0 & 0\\ \hline
Viscous & 1 & 1 & 0 & 0 & 1 & 0 \\ \hline
Almost-Uniform & 0 & 0 & 1 & 1 & 0 & 0\\ \hline
Geometric & 1 & 1 & 0 & 0 & 0 & fitted\\ \hline
\end{tabular}
    }
    \caption{\textbf{The parameter choices for different quasiconformal flow models.} The Almost-Conformal model ($A_1 = 1$, and all other parameters zero) generates a fit to the flow that is as conformal as possible. The Viscous model ($A_1 = B_1 = A_3 = 1$, and all other parameters zero) generates a minimal distortion fit to the flow. The Almost-Uniform model ($A_2 = B_2 = 1$, all other parameters zero) minimises the spatial variation of the flow in space. The Geometric model ($A_1 = B_1 = 1, C_g = 0.1,$ solve for $\lambda_1, \lambda_2, \lambda_3 \neq 0$, and all other parameters zero) fits the flow to a dynamical equation, which is a geometric flow in the example considered in this paper. }
    \label{tab:parameter}
\end{table}

\section{Computational procedure} \label{sec:discrete}
\begin{figure}[t]
 \includegraphics[width=\textwidth]{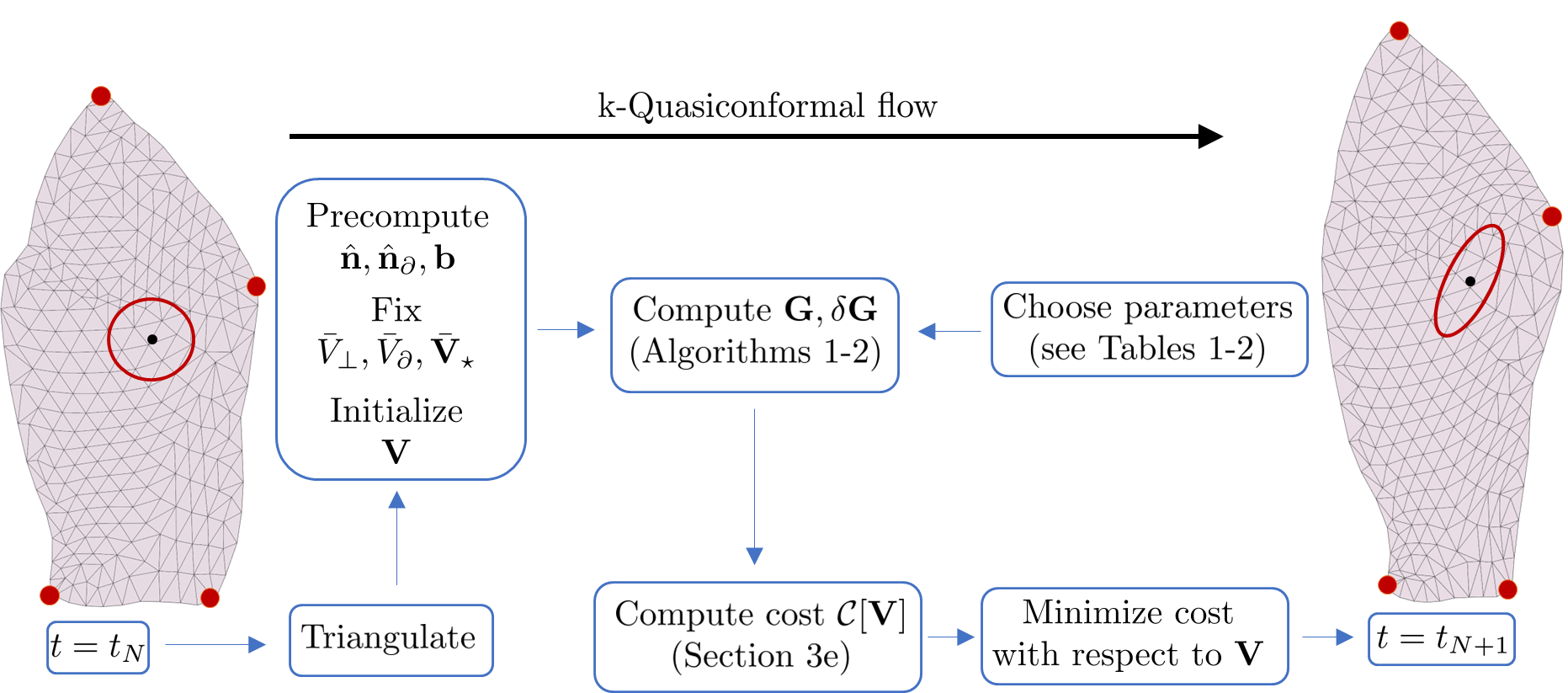}
 	\caption{\textbf{Surface registration with optimal quasiconformal flows.} This figure illustrates the workflow in our approach: We first choose a model type from those shown in Table~\ref{tab:parameter}, which gives us a fixed choice of parameters in the cost function. The flow is then broken up into discrete time steps $t_N$, triangulating each surface and precomputing geometric quantities that do not depend on the variable being optimised (the velocity field $\mathbf{V}$), such as the unit normal to the surface $\hat{\mathbf{n}}$, unit normal to the boundary curve $\hat{\mathbf{n}}_{\partial}$, and the curvature tensor $\mathbf{b}$. The component of the velocity normal to the surface $\bar{V}_{\perp}$ and boundary curve $\bar{V}_{\partial}$, in addition to the velocity of the landmark points $\bar{\mathbf{V}}_{\star}$ are fixed from the input data (see also Fig.~\ref{fig:discrete-geometry} for an illustration of these quantities). The velocity $\mathbf{V}$ is calculated for each vertex at time $t = t_N$ before moving to the next time step.} \label{fig:discretization} 
 \end{figure}

In this section, we describe how we compute and minimise the cost function defined in Eq.~\eqref{eq:total-cost-function}. We start by preprocessing input data and determining mesh connectivity and curvature, which are quantities that do not vary throughout the minimization process. We then compute the registration-dependent quantities, which are the various strain-rate measures characterising the rate of distortion of the shape data (over time and across space). Lastly, we will describe how we minimise the cost function to obtain an optimal velocity field and post-processed the results (see Fig.~\ref{fig:discretization}). 

\subsection{Data preparation and mesh preprocessing} \label{sec:preprocessing}
In this section, we describe how we pre-processed the various datasets to obtain a sequence of triangulations $\mathcal{T}(t_N) \equiv \{\mathcal{F}, \mathcal{E}, \mathcal{V} (t_N)\}$ at (equally spaced) time points $t_N$, with $1 \leq N \leq N_{max} = 30$. The connectivity of each triangulation, which is fixed over time, is defined by the faces $f_M \in \mathcal{F}$ (a list of triplets of indices, where $1 \leq M \leq M_{max}$) and edges $e_L \in \mathcal{E}$ (a list of pairs of indices, where $1 \leq L \leq L_{max}$). The list of vertex positions, which vary across time, is denoted as $\mathcal{V} (t_N)$.  

To study the growth of 2D insect wings, we considered the \textit{Manduca sexta} (tobacco hawk moth) and \textit{Junonia coenia} (buckeye butterfly) wing data from~\cite{nijhout2014development}. For each species, 2D triangular meshes were constructed based on the wing images at the larval, prepupal, pupal, and adult stages using the \texttt{distmesh} package~\cite{persson2004simple}. An initial registration between each pair of consecutive stages was then computed using a variation of the mapping method in~\cite{choi2018planar}, where the boundary correspondence was obtained using the curvature-guided matching method in~\cite{choi2018planar} and the interior correspondence was obtained using the smooth quasiconformal mapping formulation in~\cite{choi2015flash}. After getting the initial registration between the four stages above, we then applied the Piecewise Cubic Hermite Interpolating Polynomial (\texttt{pchip}) function in MATLAB to get a flow of wing shapes with smooth growth over 30 time steps, which then served as the starting point for our subsequent analysis.
 
To study the growth of 3D plant leaves using the proposed method, we considered the \textit{persea americana} (avocado) leaf dataset from~\cite{derr2018fluttering}. Each leaf scan consists of about 100 tracked points in $\mathbb{R}^3$. To create a triangular mesh for each scan, we first used the principal component analysis (\texttt{pca}) function in MATLAB to project the points onto the 2D plane and constructed a 2D Delaunay triangulation using the MATLAB \texttt{delaunay} function, with sharp triangles at the boundary removed. The planar triangulation then induced a triangulation on the set of 3D points, resulting in a triangulated leaf mesh. Next, we further applied the Taubin smoothing~\cite{taubin1995curve} and the Loop subdivision~\cite{loop1987smooth} to obtain a smooth, refined mesh with about 400 vertices. 

\subsection{Computation of geometric quantities} \label{sec:geometric}

\begin{figure}[t!]
 \includegraphics[width=\textwidth]{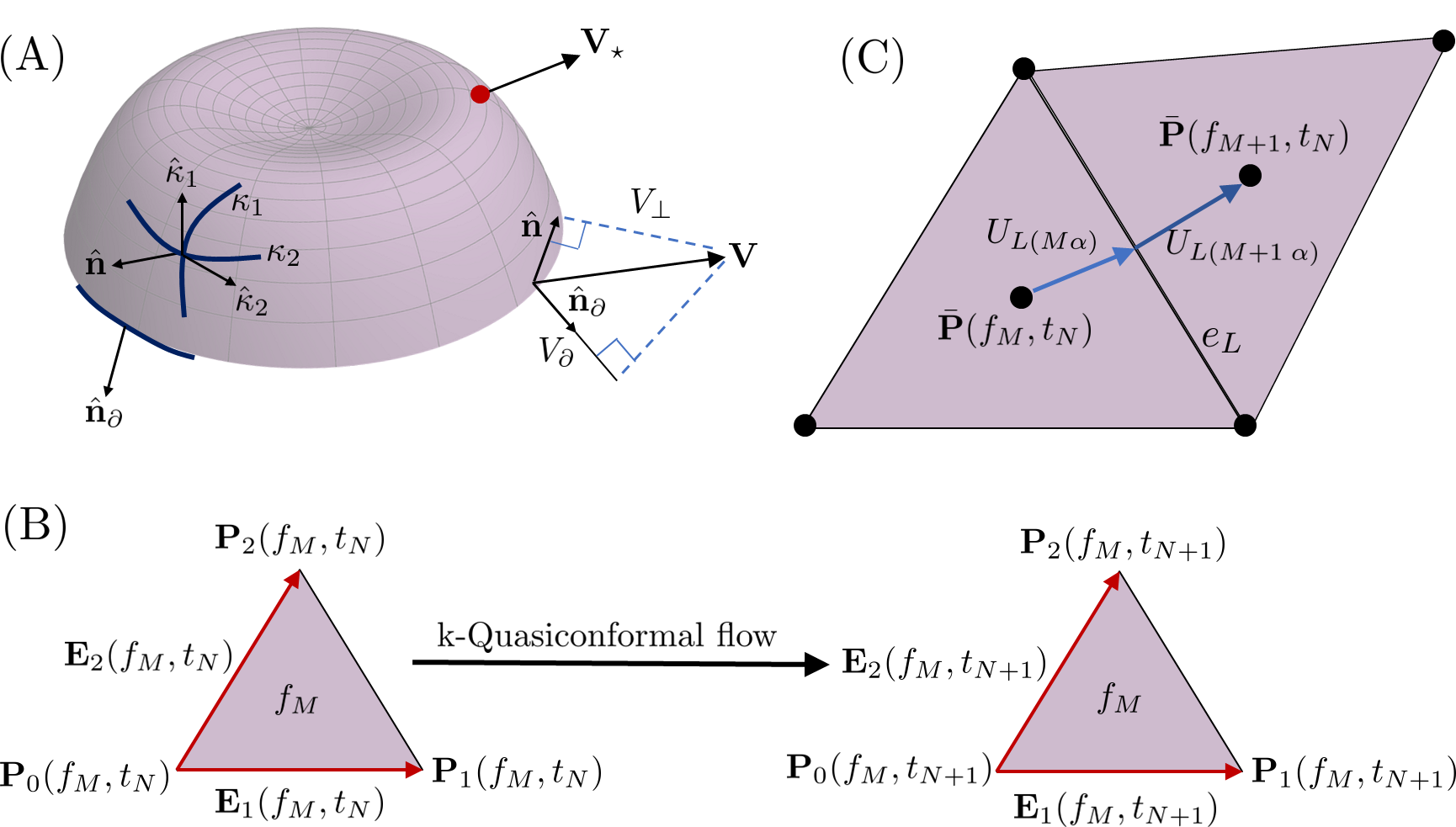}
 	\caption{\textbf{Geometric quantities used in our computational procedure.} (A) Illustrates the surface normal $\hat{\mathbf{n}}$, in addition to the principal directions ($\hat{\mathbf{\kappa}}_1, \hat{\mathbf{\kappa}}_2$) and curvatures (${{\kappa}}_1, {{\kappa}}_2$) at the same point. The vector $\hat{\mathbf{n}}_{\partial}$ is normal to both the boundary curve and $\hat{\mathbf{n}}$. The corresponding normal velocity components $V_\perp$ and $V_\partial$, defined by Eqs.~(\ref{eq:velocity}--\ref{eq:boundary-vel}), are illustrated on the lower right. The figure also illustrates the velocity $\mathbf{V}_\star$ of the landmark point shown in red. (B) Shows the vertex positions $\mathbf{P}_i(f_M, t_N)$, $i = 0,1,2$ for the triangle $f_M$ at consecutive time points $t_N$ and $t_{N+1}$. The face index satisfies $1 \leq M \leq M_{max}$ and the time index satisfies $1 \leq N \leq N_{max}$. The edge vectors $\mathbf{E}_i(f_M, t_N)$ shown in red are defined in Eq.~\eqref{eq:edge-vectors}. (C) The centroid positions of each triangle, $\bar{\mathbf{P}}(f_M, t_N)$ which are averages of the vertex positions of the triangle are indicated for two consecutive triangles $f_M$ and $f_{M+1}$. The displacement vectors starting at the centroid of triangle $M$ and ending at the midpoint of edge $e_L$ are indicated by $U_{L (M \alpha)}$, where $\alpha \in \{1,2,3\}$ indicates the component of the vector in $\mathbb{R}^3$. The parenthesis $(M \alpha)$ represents a multi-index, thus expressing $U_{L (M \alpha)}$ as an $L_{max} \times 3*M_{max}$ matrix.  } \label{fig:discrete-geometry} 
 \end{figure}

After preprocessing the meshes, we consider geometric quantities that are independent of the registration --- gauge invariant quantities as described in \cite{tumpach2015gauge} --- which allows us to pre-compute these quantities, thus reducing the time per function call during the optimization procedure. These quantities include (see Figs.~\ref{fig:discretization}-\ref{fig:discrete-geometry}) the normal vector to the surface $\hat{\mathbf{n}}$, the normal vector to the boundary curve $\hat{\mathbf{n}}_{\partial}$, the principal curvatures $\kappa_1$ and $\kappa_2$ along with the corresponding principal directions $\hat{\kappa}_1$ and $\hat{\kappa}_2$.

The boundary normal vector $\hat{\mathbf{n}}_{\partial}$ was calculated using the MATLAB \texttt{LineNormals2D} package~\cite{LineNormals2D}. For surfaces in $\mathbb{R}^3$, we computed the surface normal $\hat{\mathbf{n}}$, principal curvatures $\kappa_1, \kappa_2$, mean curvature $H = (\kappa_1+\kappa_2)/2$ and Gaussian curvature $K = \kappa_1 \kappa_2$ using the MATLAB \texttt{patchcurvature} package~\cite{patchcurvature}. The tangent and normal vectors of the boundaries of open surfaces were calculated using the MATLAB \texttt{frenet\_robust} package~\cite{frenet_robust}.

\subsection{Discretising the growth and bending strains}

In this section, we describe how we calculate the growth strain tensor Eq.~\eqref{eq:growth-strain} and bending strain Eq.~\eqref{eq:bending-cost}, as summarised in Algorithm~\ref{alg:growth-strain-tensor}.

\begin{algorithm} 
\caption{Discretise the growth and bending strains. 
  \indent \textcolor{blue}{[\textit{See Eqs.~(\ref{eq:bending-cost}--\ref{eq:growth-strain})}]}}\label{alg:growth-strain-tensor}
\begin{algorithmic}

\State \textbf{Input:} Triangulations $\mathcal{T}(t_N)$. \indent \textcolor{blue}{[ \textit{See Section~\ref{sec:preprocessing}.]}}

\State \textbf{Output:} The discretised growth $\mathbf{G}(f_M, t_N)$ and bending $\dot{\mathbf{n}}(f_M, t_N)$ strains. 

\For{$1 \leq N < N_{max}$}
\For{$1 \leq M < M_{max}$}

\State $\boldsymbol{\star}\;\;\;$ Compute basis edges $\mathbf{E}_i(f_M, t_N)$, and $\mathbf{E}_i(f_M, t_{N + 1})$.   \indent \textcolor{blue}{[  \textit{See Eq.~\eqref{eq:edge-vectors}}.]}

\State $\boldsymbol{\star}\;\;\;$ Compute the discrete metric ${\mathbf{Q}}_{\parallel}(f_M, t_N)$.   \indent \textcolor{blue}{[  \textit{See Eqs.~(\ref{eq:discrete-metric1}--\ref{eq:discrete-metric2})}.]}

\State $\boldsymbol{\star}\;\;\;$ Compute the velocity gradient, $\delta_i \textbf{V} (f_M, t_N)$. \indent \textcolor{blue}{[  \textit{See Eq.~\eqref{eq:def-gradient}}.]}

\State $\boldsymbol{\star}\;\;\;$ Compute the discretised strain rate $\dot{\mathbf{Q}}_{\parallel}(f_M, t_N)$. \indent \textcolor{blue}{[  \textit{See Eq.~\eqref{eq:discrete-strain-rate}}.]}

\State $\boldsymbol{\star\;\;\;}$ Compute the discretised growth strain tensor $\mathbf{G}(f_M, t_N)$. \indent \textcolor{blue}{[  \textit{See Eq.~\eqref{eq:discrete-growth-strain}}.]}

\State $\boldsymbol{\star\;\;\;}$ Compute the discretised bending strain $\dot{\mathbf{n}}(f_M, t_N)$. \indent \textcolor{blue}{[  \textit{See Eq.~(\ref{eq:discrete-bending-strain})}.]}

\EndFor \EndFor
\end{algorithmic}
\end{algorithm}

For a given triangle at time $t_N$, labeled with face index $f_M \in \mathcal{F}$ and spanned by the three vertices with positions $\mathbf{P}_0(f_M, t_N), \mathbf{P}_1(f_M, t_N),$ and $\mathbf{P}_2(f_M, t_N)$, the triangle centroid will approximate the smooth surface at the point $\mathbf{z} = \mathbf{z}_M$ so that (Fig.~\ref{fig:discrete-geometry}B) 
\begin{eqnarray}
    \mathbf{X}(\mathbf{z}_M, t_N) \approx \bar{\mathbf{P}}(f_M, t_N) \equiv \frac{\mathbf{P}_0(f_M, t_N)+\mathbf{P}_1(f_M, t_N) + \mathbf{P}_2(f_M, t_N)}{3}.
\end{eqnarray} 
At the same point ($\mathbf{z} = \mathbf{z}_M$), the tangent plane will be spanned by the two edge vectors (shown in red in Fig.~\ref{fig:discrete-geometry}C):
\begin{eqnarray}
    \mathbf{E}_1(f_M, t_N) \equiv \mathbf{P}_1(f_M, t_N) - \mathbf{P}_0(f_M, t_N), \;\;\;\;\;\;\; \mathbf{E}_2(f_M, t_N) \equiv \mathbf{P}_2(f_M, t_N) - \mathbf{P}_0(f_M, t_N), \label{eq:edge-vectors}
\end{eqnarray} 
which point along the edges of the triangle and correspond to the tangent vectors, $\partial_i \mathbf{X}(\mathbf{z}_M, t_N) \sim \mathbf{E}_i(f_M, t_N)$\footnote{This will be an (approximate) equality in local coordinate system $\mathbf{z}$ that parameterised a point on the triangle using $\mathbf{X}(\mathbf{z}_M,t_N) = \mathbf{P}_0(f_M, t_N) + z_1 \mathbf{E}_1(f_M, t_N)+ z_2 \mathbf{E}_2(f_M, t_N)$, where $0 \leq z_1, z_2 \leq 1$ and $z_1 + z_2 \leq 1$}. 

In order to compute the strain rate, we approximate the velocities of the vertices of the triangle (labeled by index $i = 0,1,2$), along with its centroid velocity, by 
\begin{eqnarray}
    \mathbf{V}_i(f_M, t_N) &\equiv& \frac{\mathbf{P}_i(f_M,t_{N + 1}) - \mathbf{P}_i(f_M,t_N)}{t_{N+1} - t_{N}}, \\ 
    \bar{\mathbf{V}}(f_M, t_N) &\equiv& \frac{\mathbf{V}_0(f_M, t_N) + \mathbf{V}_1(f_M, t_N) + \mathbf{V}_2(f_M, t_N)}{3}, \;\;\;
\end{eqnarray}
where the second equation defined the centroid velocity for face $f_M$. 
 To calculate the strain rate tensor, we need the spatial gradient of the velocity field, $\partial_i \textbf{V}(\mathbf{z}_M, t_N)$, which we approximate by its finite difference analogue,
 \begin{eqnarray}
     \partial_i \textbf{V}(\mathbf{z}_M, t_N) \sim \delta_i \textbf{V} (f_M, t_N) \equiv \frac{\mathbf{E}_i(f_M, t_{N+1}) - \mathbf{E}_i(f_M, t_N)}{t_{N+1} - t_{N}}, \;\;\;\;\;\;\;\;\; i = 1, 2, \label{eq:def-gradient}
 \end{eqnarray}
 where $\delta_i$ indicates a spatial finite difference. While the two vectors $\mathbf{E}_1(f_M, t_N)$ and $\mathbf{E}_2(f_M, t_N)$ span the plane tangent to the triangle labeled by index $f_M$, it will be convenient as in the continuum case to introduce the dual basis, denoted as $\mathbf{E}^{i}(f_M, t_N) \sim \partial^i \mathbf{X}(\mathbf{z}_M, t_N)$, which are defined as the (row) vectors in $\mathbb{R}^3$ satisfying
\begin{eqnarray}
    \mathbf{E}^{i}(f_M, t_N) \cdot \mathbf{E}_j(f_M, t_N) = \delta^i_j, \;\;\;\;\;\;\;\; \mathbf{E}_i(f_M, t_N) \otimes \mathbf{E}^{i}(f_M, t_N) = \mathbf{Q}_{\parallel}(f_M, t_N), \label{eq:discrete-metric1}
\end{eqnarray}
 where $\otimes$ gives a direct product between the two vectors in $\mathbb{R}^3$ and $\mathbf{Q}_{\parallel}(f_M, t_N)$ is a ($3 \times 3$) projection matrix onto the plane of the triangle indexed by $f_M$ and represents an approximation of the metric tensor in the orthonormal frame spanning the plane of the triangle: 
 \begin{eqnarray}
     \boldsymbol{\mathcal{T}}^{ij}(\mathbf{z}_M, t_N)\; g_{ij}(\mathbf{z}_M, t_N) \approx \mathbf{Q}_{\parallel}(f_M, t_N), \;\;\;\boldsymbol{\mathcal{T}}^{ij}\;\equiv\left[\partial^i  \mathbf{X}(\mathbf{z}_M, t_N) \otimes \partial^j \mathbf{X}(\mathbf{z}_M, t_N)\right]. \label{eq:discrete-metric2}
 \end{eqnarray}
 Here the factor $\boldsymbol{\mathcal{T}}^{ij}(\mathbf{z}_M, t_N)$ transforms rank-two tensors from the local coordinate basis corresponding to the (arbitrary) coordinate system $\mathbf{z}$ to a tensor in the embedding Cartesian space, $\mathbb{R}^3$ --- this will turn out to be very useful when computing the gradient cost in the next section. The fact that the two expressions for $\mathbf{Q}_{\parallel}(f_M, t_N)$ given in Eqs.~(\ref{eq:discrete-metric1}--\ref{eq:discrete-metric2}) are the same can be checked by multiplying $\mathbf{Q}_{\parallel}(f_M, t_N)$ with an arbitrary vector (field) in $\mathbb{R}^3$ and showing that in both cases it functions as a projection operator onto the tangent plane to the surface (or triangle in the discrete setting).
 
Using Eqs.~(\ref{eq:def-gradient}--\ref{eq:discrete-metric2}) and Eq.~\eqref{eq:strain-rate} we define the discrete strain rate using the time derivative (or finite difference) of ${\mathbf{Q}}_{\parallel}$, 
 \begin{equation}
 \begin{split}
        &\boldsymbol{\mathcal{T}}^{ij}(\mathbf{z}_M, t_N)\; \dot{g}_{ij}(\mathbf{z}_M, t_N) \approx  \dot{\mathbf{Q}}_{\parallel}(f_M, t_N)  \equiv  \\ 
        & {\mathbf{Q}}_{\parallel}(f_M, t_N) 
 \left[\delta_i\textbf{V}(f_M, t_N) \otimes \mathbf{E}^{i}(f_M, t_N)  +  \mathbf{E}^{i}(f_M, t_N)  \otimes \delta_i\textbf{V}(f_M, t_N) \right] {\mathbf{Q}}_{\parallel}(f_M, t_N),\label{eq:discrete-strain-rate}
 \end{split}
 \end{equation}
where the approximate equality follows by evaluating both expressions on the left and right of $\dot{\mathbf{Q}}_{\parallel}(f_M, t_N)$ --- which is a scalar with respect to the coordinate transformations on the surface and therefore can be evaluated in any coordinate system without loss of generality --- in the particular coordinate system that parameterises the triangle as $\mathbf{X}(\mathbf{z}_M,t_N) = \mathbf{P}_0(f_M, t_N) + z_1\; \mathbf{E}_1(f_M, t_N)+ z_2\; \mathbf{E}_2(f_M, t_N)$. Note that $\dot{\mathbf{Q}}_{\parallel}(f_M, t_N)$ is the \textit{projected} (onto the plane of the triangle) time derivative of ${\mathbf{Q}}_{\parallel}(f_M, t_N)$: 
\begin{eqnarray}
    \dot{\mathbf{Q}}_{\parallel}(f_M, t_N) = {\mathbf{Q}}_{\parallel}(f_M, t_N) \frac{\partial {\mathbf{Q}}_{\parallel}(f_M, t_N) }{\partial t}{\mathbf{Q}}_{\parallel}(f_M, t_N).
\end{eqnarray}

%\subsubsection{Discretising the growth strain tensor}
The last ingredient we need to compute the growth strain, given in Eq.~\eqref{eq:growth-strain}, will be the curvature tensor. Using the previously computed principal curvatures and directions, the curvature tensor is written as
 \begin{eqnarray}
      \boldsymbol{\mathcal{T}}^{ij}(\mathbf{z}_M, t_N)\; b_{ij}(\mathbf{z}_M, t_N) \approx \mathbf{B}(f_M, t_N) \equiv \kappa_1 \; \hat{\boldsymbol{\kappa}}_1 \otimes \hat{\boldsymbol{\kappa}}_1 + \kappa_2  \; \hat{\boldsymbol{\kappa}}_2 \otimes \hat{\boldsymbol{\kappa}}_2, \label{eq:discrete-curvature-tensor}
 \end{eqnarray}
where the dependence of the principal curvatures and directions on $(f_M, t_N)$ has been suppressed to simplify the expression. The equivalence of the left and right hand sides of Eq.~\eqref{eq:discrete-curvature-tensor} follows since both expressions describe a symmetric matrix whose eigenvalues are $(0, \kappa_1, \kappa_2)$ in the normal and two principal directions respectively.  

 Next, we can calculate the discrete growth strain, which following Eq.~\eqref{eq:growth-strain} is given by
 \begin{equation}
 \begin{split}
    &\boldsymbol{\mathcal{T}}^{ij}(\mathbf{z}_M, t_N)\; G_{ij}(\mathbf{z}_M, t_N) \approx \mathbf{G}(f_M, t_N) \equiv \\
    & \dot{\mathbf{Q}}_{\parallel}(f_M, t_N) - \lambda_1 \;{\mathbf{Q}}_{\parallel}(f_M, t_N)  - \lambda_2 \;  H(f_M, t_N) \;\mathbf{B}(f_M, t_N) - \lambda_3  \; K(f,t_N) \mathbf{Q}_{\parallel}(f_M, t_N). \label{eq:discrete-growth-strain}
\end{split}
 \end{equation}

In addition to the growth strain, we will need the bending strain, which is the time derivative of the normal vector used in Eq.~\eqref{eq:bending-cost}. We calculate this using the (finite) difference between the normal vector to each triangle at two consecutive time points: 
\begin{eqnarray}
\hat{\mathbf{n}}(f_M, t_N) \equiv \frac{\mathbf{E}_{1}(f_M, t_{N}) \times \mathbf{E}_{2}(f_M, t_{N}) }{\norm{\mathbf{E}_{1}(f_M, t_{N}) \times \mathbf{E}_{2}(f_M, t_{N})}}, \;\;\;\; \dot{\mathbf{n}}(f_M, t_N) \equiv \frac{\hat{\mathbf{n}}(f_M, t_{N + 1}) - \hat{\mathbf{n}}(f_M, t_N)}{t_{N+1} - t_{N}}.  \label{eq:discrete-bending-strain}
\end{eqnarray}
The growth and bending strains calculated in this section will be used in computing the total cost below, which also includes the gradient (Dirichlet energy) contributions that we turn to next.

\subsection{Discretising the growth strain gradient}

Having calculated the growth strain tensors $\boldsymbol{G}(f_M, t_N)$ in Eq.~\eqref{eq:discrete-growth-strain}, we proceed to estimate the discrete gradient, denoted as $\delta_{\alpha}\boldsymbol{G}(f_M, t_N)$, where $\alpha \in \{1,2,3\}$ indicates the direction of the derivative in $\mathbb{R}^3$. Our strategy will be to compute the gradient of $\boldsymbol{G}(f_M, t_N)$ as tensors in the embedding space ($\mathbb{R}^3$) and exploiting the relation (see SI section S1 for a derivation) between the covariant derivative of a tensor and the gradient in the Euclidean space $\mathbb{R}^3$ \cite{swarztrauber1998cartesian, ruuth-Cartesian, azencot2015discrete}. The steps are summarised in Algorithm~\ref{alg:growthstraingradient}.

\begin{algorithm}
\caption{Discretising the growth strain gradient. \indent \textcolor{blue}{[  \textit{See Eq.~\eqref{eq:grad-cost}}.]}}\label{alg:growthstraingradient}
\begin{algorithmic}
\State \textbf{Input:} Triangulations $\mathcal{T}(t_N)$ and the growth strains $\mathbf{G}(f_M, t_N)$. \ \  \textcolor{blue}{[  \textit{See Algorithm~\ref{alg:growth-strain-tensor} }.]} 

\State \textbf{Output:} The discrete growth strain $\delta \mathbf{G}(f_M, t_N)$. \indent \textcolor{blue}{[  \textit{See Eq.~\eqref{eq:discrete-grad-2}}.]}

\For{$1 \leq N < N_{max}$}

\For{all faces $f$ and all edges $e$ in the triangulation $\mathcal{T}(t_N)$}

\State $\boldsymbol{\star\;\;\;}$ Calculate the finite difference matrix ${\Delta}_{L M}$. \indent \textcolor{blue}{[\textit{See Eq.~\eqref{eq:finite-difference-relation}}.]}

\State $\boldsymbol{\star\;\;\;}$ Calculate the face-displacement vectors ${U}_{L (M\alpha)}$. \indent \textcolor{blue}{[  \textit{See Eq.~\eqref{eq:finite-difference-relation}}.]}

\State $\boldsymbol{\star\;\;\;}$ Calculate the pseudoinverse ${U}^+_{(M\alpha) L}$. \indent \textcolor{blue}{[  \textit{See Eq.~\eqref{eq:discrete-grad-0}}.]}

\State $\boldsymbol{\star\;\;\;}$ Find the gradient stiffness matrix, $\mathcal{M}_{f_1 f_2}$. \indent \textcolor{blue}{[  \textit{See Eq.~\eqref{eq:grad-matrix}}.]}

\State $\boldsymbol{\star\;\;\;}$ Find the discrete growth strain $\delta \mathbf{G}(f_M, t_N)$. \indent \textcolor{blue}{[  \textit{See Eq.~\eqref{eq:discrete-grad-2}}.]}

\EndFor \EndFor
\end{algorithmic}
\end{algorithm}

To estimate the gradient $\nabla_{\alpha} \mathbf{G}(\mathbf{z}_M, t)$, we consider a pair of triangles ($f_M$ and $f_{M+1}$ in Fig.~\ref{fig:discrete-geometry}C) connected by edge $e_L$ and denote the vector pointing from the centroid of triangle $f_M$ towards the midpoint of the edge as ${U}_{L (M\alpha)}$. On the other hand, for the second triangle in the pair, we have ${U}_{L (M+1\;\alpha)}$ will point from the midpoint of the edge $e_L$ towards its centroid, as shown in Fig.~\ref{fig:discrete-geometry}C. Lastly, when $f_M$ is a triangle not contained in the pair connected by edge $e_L$, ${U}_{L (M+1\;\alpha)} = 0$. We also define the finite difference matrix, ${\Delta}_{LM}$, which equals ${\Delta}_{L M} = -1$ for the first triangle in the pair connected by edge $e_L$, ${\Delta}_{L\; {M+1}} = 1$ for the other, and zero otherwise. Therefore, using a finite difference approximation, we have the relation
\begin{eqnarray}
    \sum_{(M\alpha)} {U}_{L (M\alpha)}\;\nabla_{\alpha}\boldsymbol{G}(f_M, t_N) \approx \sum_{M} {\Delta}_{L M} \; \boldsymbol{G}({f}_M, t_N), \label{eq:finite-difference-relation}
\end{eqnarray}
which expresses the fact that, for each edge $e_L$, dotting the gradient with the two displacement vectors connecting the triangle centroids (Fig.~\ref{fig:discrete-geometry}C) should give the difference $\boldsymbol{G}({f}_{M+1}, t_N) - \boldsymbol{G}({f}_M, t_N)$. Taking ${U}_{L (M\alpha)}$ as a matrix whose rows are labeled by the edges $L$ and columns are labeled by the multi-index $(M\alpha)$, the above linear equation can be solved to give the gradient of the growth strain tensor as
\begin{eqnarray}
    \nabla_{\alpha}\boldsymbol{G}(\mathbf{z}_M, t_N) 
 \approx \delta_{\alpha}\boldsymbol{G}(f_M, t_N) \equiv  \sum_{M_1, L} 
 {U}^+_{(M\alpha) L}\; {\Delta}_{L M_1} \; \boldsymbol{G}(f_{M_1}, t_N), \label{eq:discrete-grad-0}
\end{eqnarray}
where ${U}^+_{(M\alpha) L}$ is the pseudoinverse of the matrix ${U}_{L (M\alpha)}$. This allows us to compute the gradient squared term in the cost function using the following matrix:
\begin{eqnarray}
    \mathcal{M}^2_{M_1 M_2} \equiv \sum_{L_1, L_2, (M \alpha) } \delta \mathcal{A}_M {\Delta}_{L_1 M_1} \; {U}^+_{(M\alpha) L_1} {U}^+_{(M\alpha) L_2}{\Delta}_{L_2 M_2}, \label{eq:grad-matrix}
\end{eqnarray}
where $\delta \mathcal{A}_M$ is the area of triangle $f_M$. The matrix defined by Eq.~\eqref{eq:grad-matrix} is positive definite since it has the form of a matrix times its transpose, and therefore we may formally write its square root as ${\mathcal{M}}_{M_1 M_2}$. The advantage of introducing the matrix ${\mathcal{M}}_{M_1 M_2}$ is that it depends on the triangulation $\mathcal{T}(t_N)$, and can be precomputed so that during optimization, calculating the discrete growth strain gradient is done by the following matrix multiplication (see SI Section~S1),
\begin{eqnarray}
    \delta \boldsymbol{G}(f_{M}, t_N) \equiv \frac{1}{\delta \mathcal{A}_{M}^{1/2}}\sum_{M_1} 
 {\mathcal{M}}_{M M_1} \;\boldsymbol{G}(f_{M_1}, t_N). \label{eq:discrete-grad-2}
\end{eqnarray}
Thus, during optimization, only the growth strain $\boldsymbol{G}(f_{M}, t_N)$ needs to be computed, and the gradient term follows my matrix multiplication, which enables significantly faster implementation of the algorithm --- through vectorisation \cite{plangger2016vectorization}--- using the packages JAX and numpy in python \cite{numpy, jax}. Once obtained, the quantity $\delta \boldsymbol{G}(f_{M}, t_N)$ is used to compute the gradient contribution to the cost function in a manner similar to that used in computing the viscous term, as we show next (see also SI section S1).

\subsection{Discretising and minimising the total cost} \label{sec:discrete-total-cost}

Having calculated the growth strain along with its gradients, we now proceed to calculate the total cost function defined in Eq.~\eqref{eq:total-cost-function} by adding up all the different contributions. Using Eq.~\eqref{eq:discrete-bending-strain}, the bending contribution to the cost is 
\begin{eqnarray}
    \mathcal{C}_{bend} = A_3 \sum_{M = 1}^{M_{max}} \sum_{N = 1}^{N_{max}}\delta t_N \; \delta \mathcal{A}_M  \left[ \dot{\mathbf{n}}(f_M, t_N) \cdot \dot{\mathbf{n}}(f_M, t_N) \right], \label{eq:discrete-bend-cost}
\end{eqnarray}
where $\delta \mathcal{A}_M$ is the area of the triangle $f_M$ at time $t = t_N$, $\delta t_N \equiv t_{N+1} - t_N$, and we sum over all triangles and time steps. 
We compute the contribution of triangle $f_M$, at time $t_N$, to the viscous cost defined in Eq.~\eqref{eq:viscous-cost} through
\begin{eqnarray}
    \mathcal{C}_{viscous} = \sum_{M = 1}^{M_{max}} \sum_{N = 1}^{N_{max}}\delta t_N \; \delta \mathcal{A}_M  \left( \left(A_1 - B_1/2\right) Tr\left[\mathbf{G}(f_M, t_N)\right]^2  + B_1 Tr\left[\mathbf{G}(f_M, t_N)^2\right]\right), \label{eq:discrete-viscous-cost}
\end{eqnarray}

The gradient cost defined in Eq.~\eqref{eq:grad-cost} is calculated (see SI Section~S1) as in Eq.~\eqref{eq:discrete-viscous-cost} with $\mathbf{G}(f_M, t_N)$ replaced with $\delta \mathbf{G}(f_M, t_N)$:
\begin{eqnarray}
     \mathcal{C}_{grad} = \sum_{M = 1}^{M_{max}} \sum_{N = 1}^{N_{max}}\delta t_N \; \delta \mathcal{A}_M  \left( \left(A_2 - B_2/2\right) \;Tr\left[\delta \mathbf{G}(f_M, t_N)\right]^2  + B_2 Tr\left[\delta \mathbf{G}(f_M, t_N)^2\right]\right).\label{eq:discrete-gradient-cost}
\end{eqnarray}

The soft constraints forcing the normal velocities, defined in Eqs.~(\ref{eq:normal-cost}--\ref{eq:boundary-cost}), can be calculated as
\begin{eqnarray}
    \mathcal{C}_{normal} = C_n \sum_{M = 1}^{M_{max}} \sum_{N = 1}^{N_{max}}\delta t_N \; \delta \mathcal{A}_M \left[ {\hat{\mathbf{n}}}(f_M, t_N) \cdot \mathbf{V}(f_M, t_N) - \bar{V}_{\perp}(f_M, t_N) \right]^2, \label{eq:discrete-normal-cost}\\
    \mathcal{C}_{boundary} = C_b \sum_{L = 1}^{L_{max}} \sum_{N = 1}^{N_{max}}\delta t_N \; \delta \mathcal{L}_L  \left[ {\hat{\mathbf{n}}_{\partial}}(e_L, t_N) \cdot \mathbf{V}(e_L, t_N) - \bar{V}_{\partial}(e_L, t_N) \right]^2,   \label{eq:discrete-boundary-cost}
\end{eqnarray}
where Eq.~\eqref{eq:discrete-boundary-cost} is summed over the boundary edges $e \in \mathcal{E}_\partial$ with lengths given by $\mathcal{L}_L$.

The last contribution we must add is the landmark constraint given in Eq.~\eqref{eq:landmark-cost}
\begin{eqnarray}  
\mathcal{C}_{landmark} =  C_L  \sum_{\star} \sum_{N = 1}^{N_{max}}\delta t_N \left[\mathbf{V}_*(t_N) - \bar{\mathbf{V}}_\star(t_N)\right]^2, \;\;\;\label{eq:discrete-landmark-cost}
\end{eqnarray}
where, as before, $\star$ indexes the landmark vertices, $\mathbf{V}_*(t_N)$ is the velocity of the landmark vertex, while $\bar{\mathbf{V}}_*(t_N)$ is the prescribed landmark velocity.

Finally, we can sum up the contributions to get the total cost as in Eq.~\eqref{eq:total-cost-function},
\begin{eqnarray}
    &\mathcal{C}_{total}\left[\mathcal{T}(t_N)\right] = \mathcal{C}_{viscous} + \mathcal{C}_{grad} + \mathcal{C}_{bend}  + \mathcal{C}_{normal} + \mathcal{C}_{boundary} + \mathcal{C}_{landmark}. \;\;\;\;\;\;\label{eq:-discrete-cost-function} 
\end{eqnarray}

\subsection{Minimisation and post-processing}
\begin{figure}[t]
 \includegraphics[width=\textwidth]{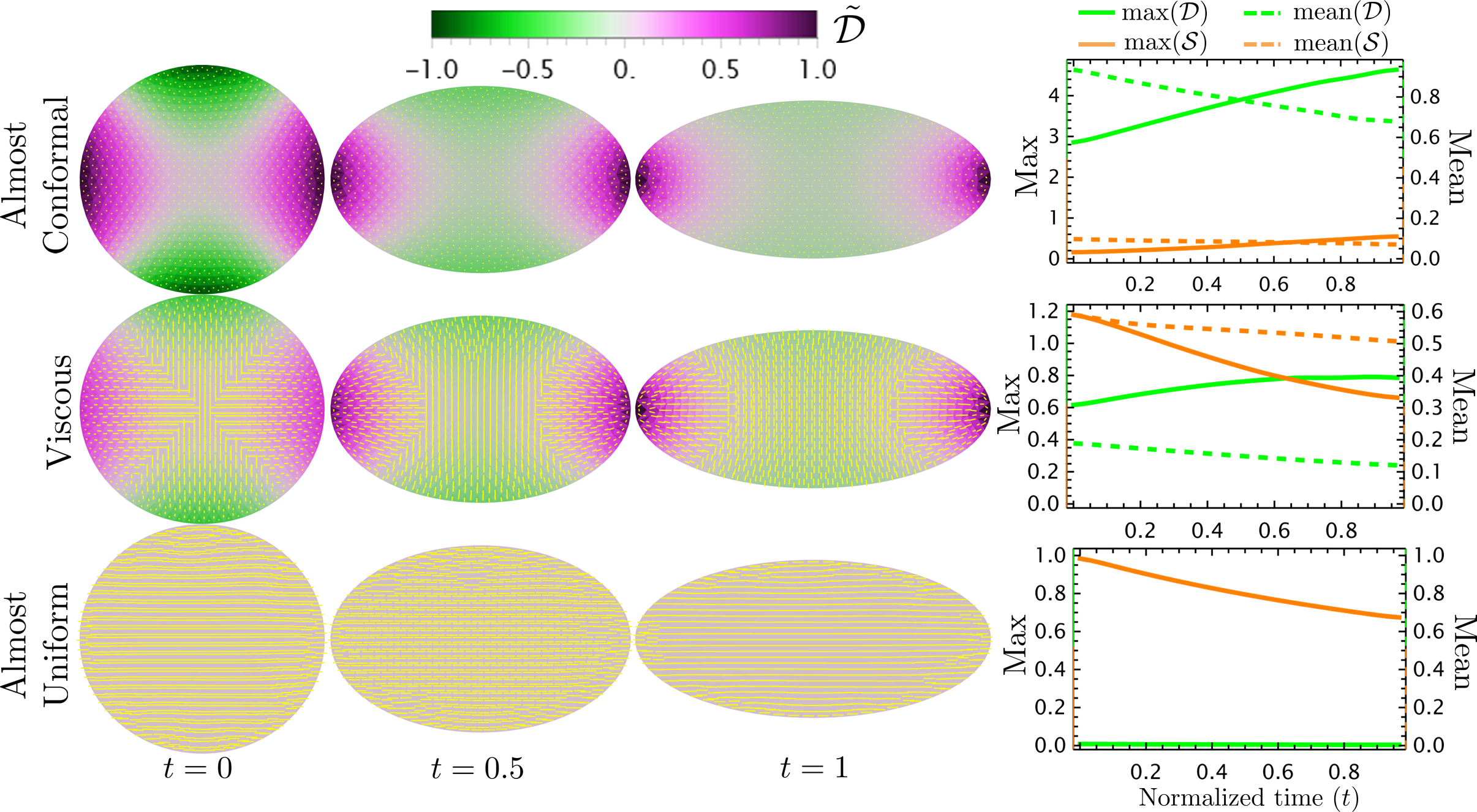}
 	\caption{\textbf{Optimal 2D quasiconformal flows for a disc undergoing simple shear.} The starting point is a disc undergoing expansion by a factor of $3/2$ in the horizontal direction, while shrinking by the same factor in the vertical, with area conserved at all times. To infer the possible growth patterns, we run the optimization with three different parameter choices given in Table~\ref{tab:parameter}: Almost-Conformal ($A_1 = 1$, all other parameters zero), Viscous ($A_1 = B_1 = A_3 = 1$, and all other parameters zero), and Almost-Uniform ($A_2 = B_2 = 1$, and all other parameters zero). Here $\mathcal{D}$ is the dilation rate given in Eq.~\eqref{eq:dilation} and $\mathcal{S}$ is the shear rate given in Eq.~\eqref{eq:shear}. The triangles are color-coded according to the normalised dilation rate, $\tilde{\mathcal{D}} \equiv \mathcal{D}/ \max(\mathcal{D}, \mathcal{S})$, while the length of the yellow line segments represents the magnitude of normalised shear rate, $\tilde{\mathcal{S}} \equiv \mathcal{S}/ \max(\mathcal{D}, \mathcal{S})$, and their direction is along the largest eigen direction of the strain rate tensor. The last column shows the maximum and mean dilation and shear rates plotted against normalised time for the different registrations in each row.  
  }  \label{fig:shear-disc} 
 \end{figure}

To speed up the computation and minimization of Eq.~\ref{eq:-discrete-cost-function}, we use the package JAX for just-in-time (JIT) compilation and automatic differentiation \cite{jax}. We use the limited memory BFGS \cite{liu1989limited} to find the optimal velocity field for each vertex at each time step $T_N$. Once an optimal velocity field is obtained, it can be used to analyze the resulting deformation fields: For each triangle $f_M$ and time $t_N$ we calculate the growth strain tensor using the velocity field as explained in the previous sections. To visualise the growth strain tensor $\mathbf{G}(f_M, t_N)$ --- that resulted from the minimization procedure --- we calculate its eigenvalues and eigenvectors. One eigenvalue will be zero, with corresponding eigenvector normal to the triangle, while the other two ($\mathcal{G}_1(f_M, t_N), \mathcal{G}_2(f_M, t_N)$) will be in the plane of the triangle. The sum of the eigenvalues will be the dilation rate, $\mathcal{D}(f_M, t_N) \equiv \mathcal{G}_1(f_M, t_N) + \mathcal{G}_2(f_M, t_N)$, while the difference gives the shear rate, $\mathcal{S}(f_M, t_N) \equiv |\mathcal{G}_1(f_M, t_N) -  \mathcal{G}_2(f_M, t_N)|$. The dilation rate $\mathcal{D}(f_M, t_N)$ is represented using a color code (see Fig.~\ref{fig:shear-disc}), while the shear rate $\mathcal{S}(f_M, t_N)$ is illustrated using the length of line segments that point along the eigenvector corresponding to the largest eigenvalue of $\mathbf{G}(f_M, t_N)$.

when post-processing our results, we perform a smoothing step by applying a Gaussian filter to the dilation and shear rates. This is done in Mathematica \cite{Mathematica} by using the function GraphDistanceMatrix that calculates the geodesic distances $d_{M_1, M_2}(t_N)$ between the centroids of triangles $f_{M_1}$ and $f_{M_2}$ in the mesh $\mathcal{T}(t_N)$ (represented as a weighted graph with edge weights given by their lengths). Smoothing of a quantity $\mathcal{Q}(f_M, t_N)$ is then given by 
\begin{eqnarray}
    \mathcal{Q}_{smooth}(f_{M_1}, t_N) = \frac{1}{\mathcal{N}(M_1)} \sum_{M_2 = 1}^{M_{max}} \exp{\left(- \frac{d^2_{M_1, M_2}(t_N)}{2 \Bar{d}^2}\right)} {Q}(f_{M_2}, t_N), 
\end{eqnarray}
where $\Bar{d}$ is the smoothing length scale of the filter, which we take as $10\%$ the maximum distance between any two triangles and $\mathcal{N}(M_1) \equiv \sum_{M_2 = 1}^{M_{max}} \exp{\left(- {d^2_{M_1, M_2}(t_N)}/{2 \Bar{d}^2}\right)}$ is the filter normalization factor. 

\section{Experimental results} \label{sec:results}
To demonstrate the effectiveness of our proposed framework, we consider different classes of surfaces starting from simple 2D shapes with and without landmarks to more complex closed or open surfaces in 3D.

\subsection{2D shapes without landmarks}
Planar shapes are a special example of surfaces with boundary that are confined to the plane. In this case, all normal displacements vanish, $\bar{V}_\perp(\mathbf{z}, t) = 0$, and the first term of Eq.~\eqref{eq:normal-cost} is minimised by setting ${V}_\perp(\mathbf{z}, t) = 0$ everywhere. We start by studying the deformation of a disc undergoing simple shear, shrinking by a factor of $1.5$ in the vertical direction and expanding by the same factor in the horizontal direction over the course of 30 discrete time steps (see Fig.~\ref{fig:shear-disc}).

We perform three different minimisations, corresponding to the almost-conformal, viscous, and almost-uniform models described in Table~\ref{tab:parameter}. The almost-conformal model minimises the shear rate (absolute difference in eigenvalues of the strain rate tensor) and will therefore find a nearly conformal fit to the flow (see Fig.~\ref{fig:shear-disc}). The Viscous model minimises a combination of shear dilation rates (sum of eigenvalues of the strain rate tensor), while the almost-uniform model favors a strain rate tensor that is uniform in space. Line segments in the third row of Fig.~\ref{fig:shear-disc} point in the same direction and have the same length, indicating a constant shear rate, while the dilation rate is zero everywhere. Note that the almost-uniform fit to the flow recovers the constant simple shear transformation that was used to generate this test example, while the almost-conformal and viscous fits predict a completely different growth pattern. Interestingly, as shown in the second row of Fig.~\ref{fig:shear-disc}, the viscous fits produce domains with distinct orientations and where the dilation rate changes from position (expansion) to negative (contraction).

\subsection{2D shapes with landmarks}

In this section, we assume that a subset of the vertices of the mesh are given as landmarks, points where a trajectory $\mathbf{x_p}(t)$ is given \textit{a priori},  (e.g., by florescent labeling). These trajectories are enforced in the minimisation procedure by adding the corresponding cost as described in Eq.~\eqref{eq:landmark-cost} and implemented as described in Eq.~\eqref{eq:discrete-landmark-cost}. We perform the same three minimizations as in the previous section (see Table~\ref{tab:parameter}).

\begin{figure}  [t!]
 \includegraphics[width=\columnwidth]{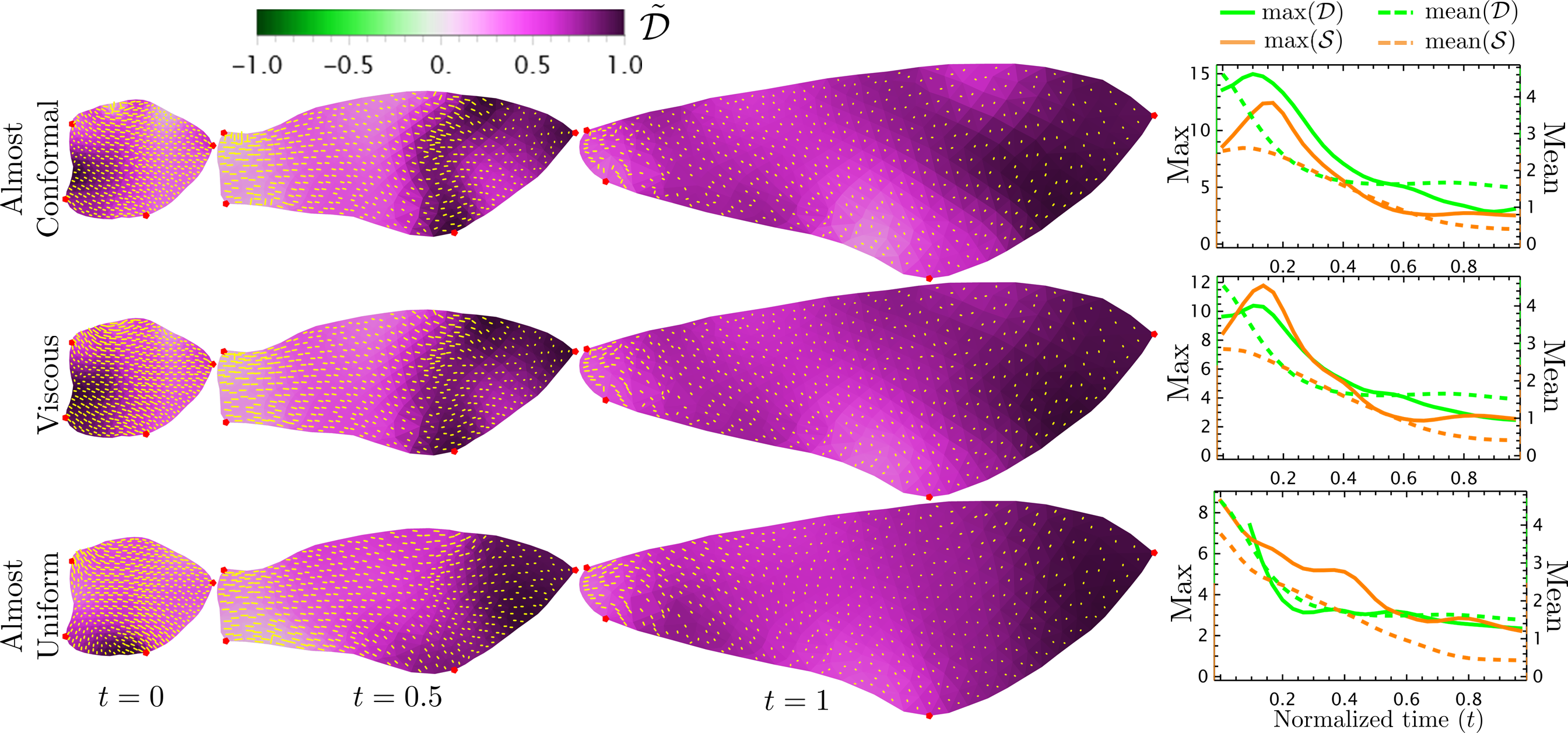}
 	\caption{\textbf{Optimal 2D quasiconformal flows for \textit{Manduca sexta} (tobacco hawk moth) wings registration with landmarks.} Images of growing \textit{Manduca sexta} wings are taken from \cite{nijhout2014development}. To infer the possible growth patterns, we run the optimization with three different parameter choices given in Table~\ref{tab:parameter}: Almost-Conformal, Viscous, and Almost-Uniform (see the caption of Fig.~\ref{fig:shear-disc} for details). See also Movie S1 for the full quasiconformal flow process. 
  } \label{fig:moth-landmarks} 
 \end{figure}

Figure~\ref{fig:moth-landmarks} shows the results of our analysis performed on \textit{Manduca sexta} (tobacco hawk moth) data, which was taken from~\cite{nijhout2014development}. For the almost-conformal model, we observe that the more landmarks we have, the less conformal --- as determined by higher values of $\mathcal{S}/\mathcal{D}$ shown in Fig.~\ref{fig:moth-landmarks} compared with SI Fig.~S1. In addition, the inferred growth patterns depend on the assumed model, just as in the disc example (Fig.~\ref{fig:shear-disc}). For example, in the absence of landmark constraints (SI Fig.~S1), the almost-conformal model contains area shrinking (green regions) and expansion (purple regions), while the viscous and almost-uniform models contain only area expansion. Interestingly, we observe that the proximal and distal parts of the wing behave differently in all models with a wave of expansion moving from the proximal to the distal direction over time (see SI Movie~S1). 

\begin{figure}  [t!]
 \includegraphics[width=\columnwidth]{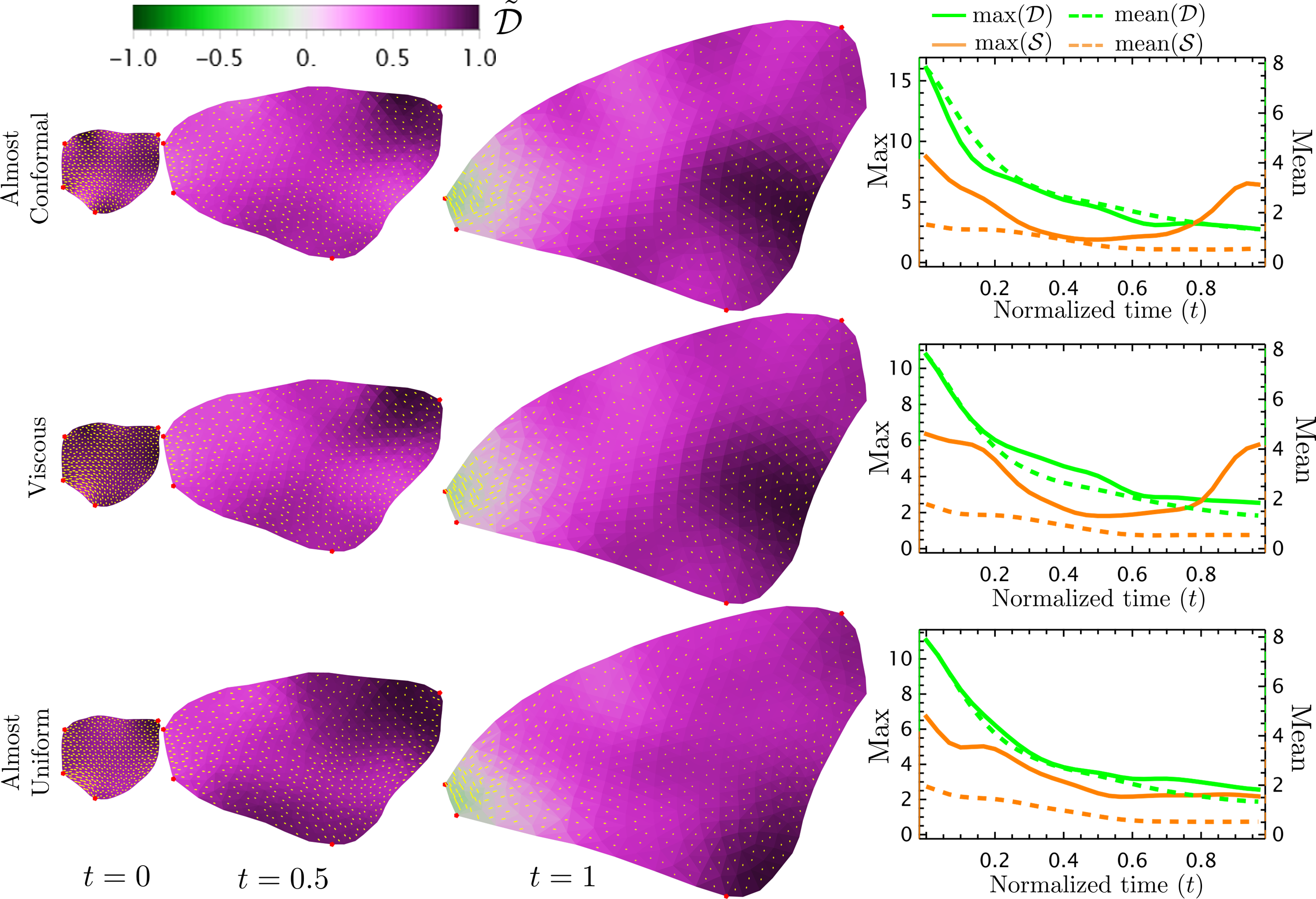}
 	\caption{\textbf{Optimal 2D quasiconformal flows for \textit{Junonia coenia} (buckeye butterfly) wing registration with landmarks.} Images of growing \textit{Junonia coenia} wings are taken from \cite{nijhout2014development}. To infer the possible growth patterns, we run the optimization with three different parameter choices given in Table~\ref{tab:parameter}: Almost-Conformal, Viscous, and Almost-Uniform (see the caption of Fig.~\ref{fig:shear-disc} for details). See also SI Movie~S2 for the full quasiconformal flow process. 
  } \label{fig:butterfly-landmarks} 
 \end{figure}

The corresponding results for \textit{Junonia coenia} (buckeye butterfly) wing data, also taken from~\cite{nijhout2014development}, are shown in Fig.~\ref{fig:butterfly-landmarks} (see also SI Fig.~S2 and SI Movie~S2). In this case, the transformation appears to be more conformal (isotropic) in nature, with the viscous and almost-uniform models exhibiting higher dilation rates compared with shear rates.

\begin{figure}  [t]
 \includegraphics[width=\columnwidth]{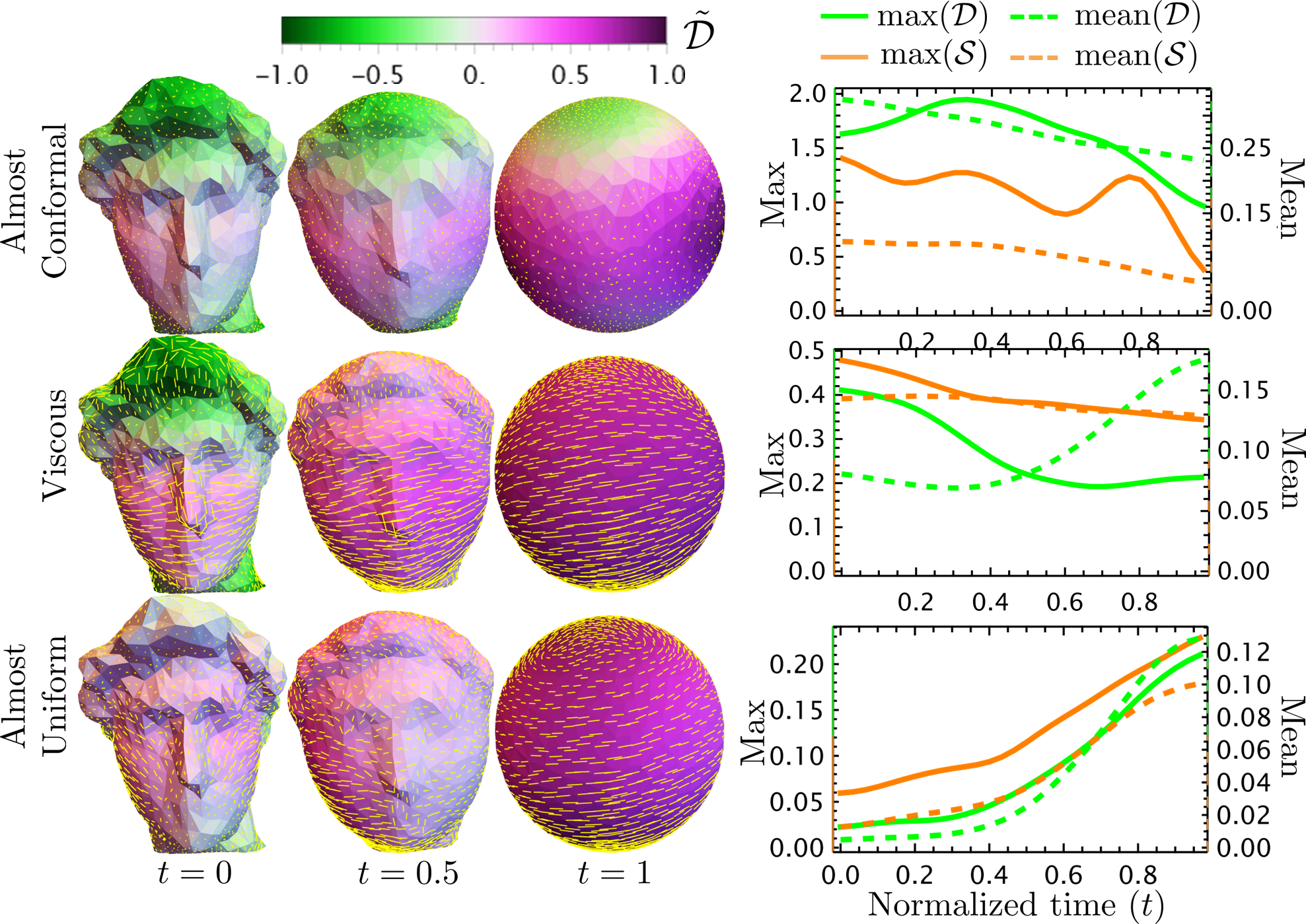}
 	\caption{\textbf{Optimal 3D quasiconformal flows for closed surfaces.} Here we consider a head mesh evolving towards a sphere. We run the optimization with three different parameter choices given in Table~\ref{tab:parameter}: Almost-Conformal, Viscous, and Almost-Uniform (see the caption of Fig.~\ref{fig:shear-disc} for details). see also SI Movie~S3 for the full quasiconformal flow process. 
  } \label{fig:david} 
 \end{figure}

\subsection{Closed surfaces in 3D}

For surfaces embedded in 3D, vertices may move in all three directions, but their movement is constrained by the contributions to the cost function given in Eq.~(\ref{eq:normal-cost}--\ref{eq:boundary-cost}), which enforce the displacement of each vertex to be parallel to the surface and the boundary curve. For closed surfaces (our analysis should work independent of the topology of the surface as long as the flow is smooth), we do not need to enforce boundary displacements and we may run the same three registration models as before (almost-conformal, viscous, almost-uniform). The first example (SI Fig.~S3) is a sphere expanding in one direction by a factor of 1.5 and contracting in the other two to preserve volume in the course of 30 times steps. We see that the conformal model (first row in SI Fig.~S3) generates nearly conformal flows (small shear rates), where expansion is initially concentrated at the poles of the sphere while contraction dominates in the equator. 

The second example we consider (see Fig.~\ref{fig:david} and SI Movie S3) is a head deforming into a sphere with a linear interpolation in the course of 30 time steps. Unlike the first example (SI Fig.~S3), the almost-conformal model when deforming the head does not reduce the shear rate to zero (orange curve in the first row of Fig.~\ref{fig:david}), which happens at sharp features on the mesh, where it becomes harder to estimate the normal vector and tangent plane to the surface. On the other hand, the uniform model leads to a smoother pattern of dilation and shear rates (both in direction and magnitude), including at sharp features.

\subsection{Surfaces with boundary in 3D}

\begin{figure}  [t!]
 \includegraphics[width=\columnwidth]{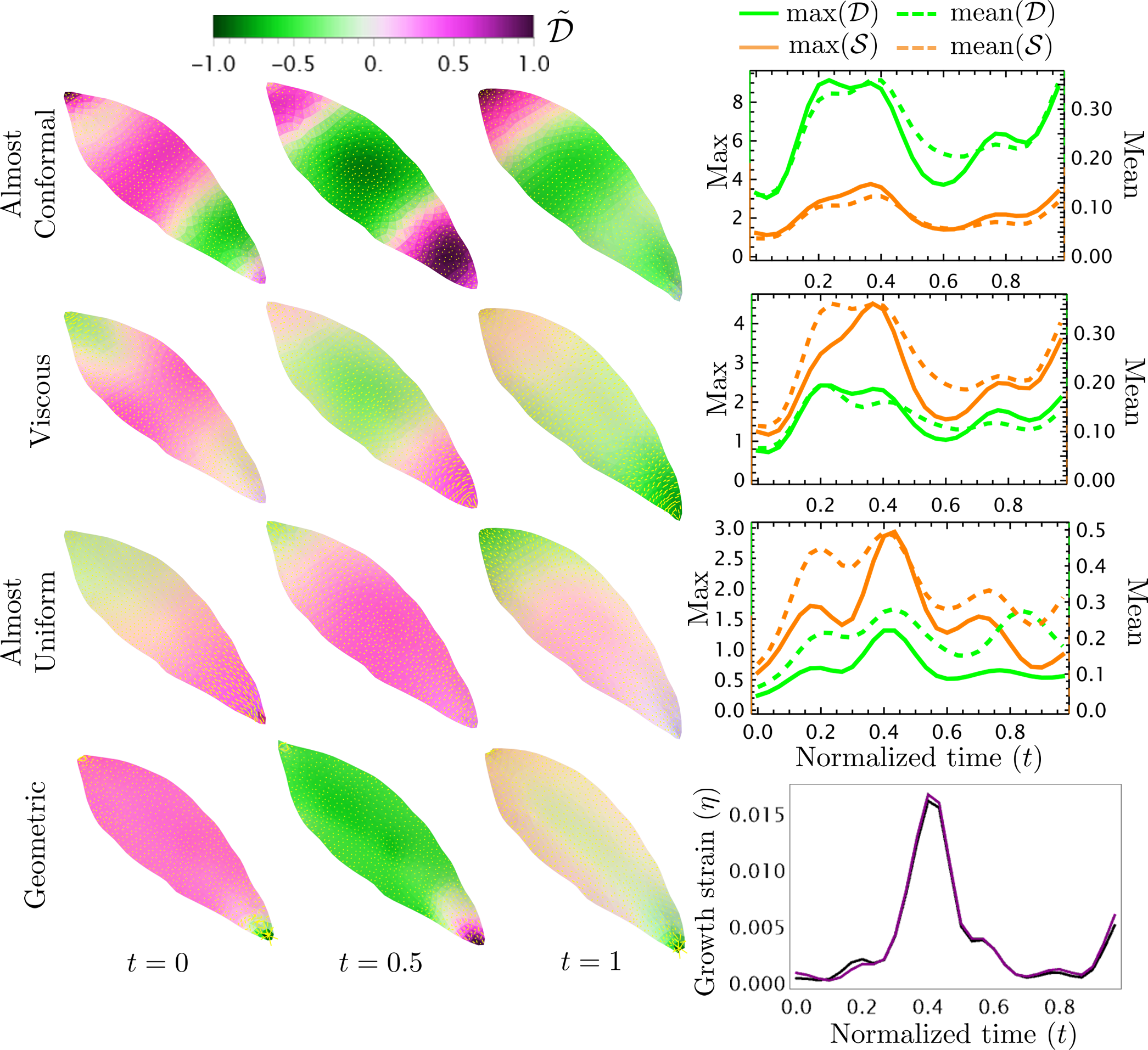}
 	\caption{\textbf{Optimal 3D quasiconformal flows using geometric flows for growing leaves.} Leaf data was taken from \cite{derr2018fluttering} and processed as described in the text and Fig.~\ref{fig:discretization}. We run the optimization with four different parameter choices given in Table~\ref{tab:parameter}: Almost-Conformal, Viscous, Almost-Uniform, and Geometric (see the caption of Fig.~\ref{fig:shear-disc} for details). See also SI Movie S4 for the full quasiconformal flow process. The plot on the lower right gives the magnitude of the growth cost, which is defined as $\eta(t) \equiv  \langle \mathcal{D}(\mathbf{z}, t) + \mathcal{S}(\mathbf{z}, t) \rangle/2$, where the angle brackets represent averaging over the area of the surface. The purple curve represents the viscous model, while black represents the geometric model. 
  } \label{fig:leaf-growth} 
 \end{figure}

We test our algorithm on a \textit{persea americana} (avocado) leaf dataset~\cite{derr2018fluttering}, using almost-conformal, viscous, and almost-uniform models (Table~\ref{tab:parameter}). We see from the results shown in Fig.~\ref{fig:leaf-growth} (see also SI Movie~S4), that the conformal model leads to low values of the mean shear rate, and therefore a nearly conformal flow. However, it still leads to nonzero shear rate near the tips of the leaf, where the margin deviates from a smooth curve. As before, the inferred growth pattern depends strongly on the model assumed (almost-conformal, viscous, or almost-uniform). Therefore, we now explore how modeling the dynamics generating a given flow can help us better recover the growth patterns.

To validate our algorithm, we generate a synthetic data set involving a cylinder deforming according to the Ricci flow --- a solution to $G_{ij} = 0$, where $G_{ij}$ is given in Eq.~\eqref{eq:growth-strain} and $\lambda_1 = \lambda_2 = 0$, and $\lambda_3 = 0.007$ --- as given by the following equation (illustrated in SI Fig.~S5), 
\begin{eqnarray}
    r(z_3) = \left(r_0 +  r_1 \sin(k z_3) e^{\lambda t} \right) \hat{r}(\phi) + \left(z_3 - \frac{r_1}{r_0 k} \cos(k z_3) 
    \left[e^{\lambda t} - 1\right] \right) L_0 \hat{z}_3. \label{eq:ricci-cyl}
\end{eqnarray}
Here $z_3 \in [0,1]$ is the vertical coordinate increasing along the length of the cylinder, $\phi \in [0,2 \pi]$ is the azimuthal coordinate, $\hat{r}$ and $\hat{z}_3$ are cylindrical basis vectors, $\lambda = 1$ is a growth rate (multiplied by the duration of the flow $T$), $r_0 = 0.16$ is the unperturbed cylinder radius, $r_1 = 0.1 r_0$ is the initial amplitude of the deformation, and $L_0 = 1$ is the initial height of the cylinder. 

To determine whether the dynamical equation $G_{ij} = 0$ explains the given flow of surfaces $\mathbf{X}(\mathbf{z}, t)$, we minimise the cost function in Eq.~\eqref{eq:total-cost-function} using the Geometric model parameters given in the last row of Table~\ref{tab:parameter}, where $\lambda_1, \lambda_2, \lambda_3 \neq 0$, and comparing it to a Viscous model (having the same $A_1, B_1$ but with $\lambda_1 = \lambda_2 =\lambda_3 = 0$). In addition, we perform separate fits for the Ricci ($\lambda_1 = \lambda_2 = 0, \lambda_3 \neq 0$) and Mean ($\lambda_1 = \lambda_3 = 0, \lambda_2 \neq 0$) curvature flows. We find that, as expected, the Geometric and Ricci fits, when applied to the growing cylinder data set, significantly reduce the growth cost (SI Fig.~S4--S5), especially for early times, as expected since the data set was generated using Eq.~\eqref{eq:ricci-cyl}, which is an approximate solution to the Ricci flow, assuming small deviations from a cylinder. Furthermore, the Ricci fit recovers the value $\lambda_3 \approx 0.007$ (with less than $3\%$ error). 

The last row of Fig.~\ref{fig:leaf-growth} shows the results for a growing leaf. While we obtain a correlation between curvature and dilation rate, the minimum growth cost $\mathcal{C}_{viscous}$ is not significantly reduced in the Geometric model. This implies that while growth rate may be correlated with curvature in leaves, the equation $G_{ij} = 0$ is not sufficient to explain the entire growth pattern, some of which may be prescribed in-plane growth that is not curvature dependent or is triggered by movement of the mid-vain as described in \cite{derr2018fluttering}.

\section{Conclusions}\label{sect:conclusion}
In this paper, we have extended the notion of $K$-quasiconformal maps to $k$-quasiconformal flows for modeling the continuous growth process of biological shapes. 

From a biological perspective, understanding the growth and form of shapes is a central problem in biology. The flexibility of our proposed framework allows us to infer growth patterns using a wide range of models with different properties, thereby paving a new way for unveiling the growth dynamics of different biological structures. In addition, besides the almost-conformal, viscous, almost-uniform, and geometric models considered in our current framework, we plan to further consider other dynamical laws and test them using our framework, for example those that involve nonlocal feedback between mechanical stress and growth \cite{al2022grow,harline2022dynamic}.

From a mathematical perspective, it is worth noting that an optimal quasiconformal flow between two shapes generally differs from a linear temporal interpolation of an optimal quasiconformal map between two shapes. Therefore, the use of quasiconformal flows is important to capture the spatio-temporal variation of the shear and dilatation fields during the growth process and opens new mathematical questions for future work.

From an algorithmic perspective, we have proposed a unifying framework for inferring growth patterns using a diverse set of criteria, including local geometric distortions, spatial variations, and fits to dynamical equations, such as geometric flows. Specifically, we can effectively fit different models to a sequence of observed data by considering different combinations of the cost functions described in our work. As demonstrated by our experimental results on different 2D and 3D shapes, including insect wings, plant leaves, and other open or closed surfaces, our framework is useful whenever a growing structure --- in nature or industry --- is observed but the identities of individual points are not tracked over time (except possibly for a small set of landmarks). 

Our approach depends on minimal finite difference approximations --- for example see Eqs.~(\ref{eq:def-gradient}--\ref{eq:discrete-strain-rate}) --- valid up to first order in $\delta t_N$, and is therefore best suited to situations with dense sampling of the growth process. The cost function Eq.~\eqref{eq:total-cost-function} can be thought of as a Riemannian metric  \cite{klassen2004analysis} that gives lengths of paths (or flows) in the space of parameterised surfaces. Therefore, extending our methods beyond first order approximations in time amounts to higher order interpolations of these geodesics. While such geodesics flows have been calculated in the past for special cost functions \cite{kurtek2011elastic}, the advantage of the current approach is greater flexibility in choosing the cost function. 

Analogously, in the spatial domain, our computations rely on decomposing the displacements into components that are tangent and normal to the surface and its boundary. Such decomposition may not be well-behaved for irregular meshes with sharp features. These potential limitations may be overcome by following the philosophy of discrete differential geometry, aiming to develop methods robust for discrete meshes \cite{grinspun2006discrete, crane2017discrete, crane2018discrete}. Therefore, it will be fruitful to define discrete quasi-conformal flows in future iterations. 

Another direction for future work is the comparison of hard and soft constraints to enforce landmark and normal displacements that are taken from the input data. The advantages of soft constraints are ease of implementation and tunability of the relative importance of input data compared with model prediction. This is analogous to optimal filters used in control theory to tune the relative importance of measurement data vs model predictions based on the strength of the noise from each source (measurement uncertainty and model error), or equivalently, the role of priors and data in the context of inference problems. More generally, there is much to be gained by combining the power of geometrical approaches that harness the structured information in morphodynamics with data-driven computational models that are efficient and fast, and this paper takes a step in this direction.

\bibliographystyle{ieeetr}
\bibliography{reference}

\end{document}